\documentclass[numberedappendix,natbib209]{emulateapj}

\usepackage{epsfig}
\usepackage{longtable}
\usepackage{natbib}

\begin{document}

\title{A Public K$_{s}$-selected Catalog in the COSMOS/UltraVISTA Field: Photometry, Photometric Redshifts and Stellar Population Parameters\altaffilmark{1,2}}
\author{Adam Muzzin\altaffilmark{3}, Danilo Marchesini\altaffilmark{4}, Mauro Stefanon\altaffilmark{5}, Marijn Franx\altaffilmark{3}, Bo Milvang-Jensen\altaffilmark{6}, James S. Dunlop\altaffilmark{7}, J. P. U. Fynbo\altaffilmark{6}, Gabriel Brammer\altaffilmark{8}, Ivo Labb\'{e}\altaffilmark{3},  Pieter van Dokkum\altaffilmark{9}}
\altaffiltext{1}{Based on data products from observations made with ESO Telescopes at the La Silla Paranal Observatory under ESO programme ID 179.A-2005 and on data products produced by TERAPIX and the Cambridge Astronomy Survey Unit on behalf of the UltraVISTA consortium.}
\altaffiltext{2}{Catalog and other data products are available at http://www.strw.leidenuniv.nl/galaxyevolution/ULTRAVISTA/}
\altaffiltext{3}{Leiden Observatory, Leiden University, PO Box 9513,
  2300 RA Leiden, The Netherlands}
\altaffiltext{4}{Department of Physics and Astronomy, Tufts University, Medford, MA 06520, USA}
\altaffiltext{5}{Physics and Astronomy Department, University of Missouri, Columbia, MO 65211}
\altaffiltext{6}{Dark Cosmology Centre, Niels Bohr Institute, University of Copenhagen, Juliane Maries Vej 30, 2100 Copenhagen, Denmark}
\altaffiltext{7}{SUPA, Institute for Astronomy, University of Edinburgh, Royal Observatory, Edinburgh EH9 3HJ, UK}
\altaffiltext{8}{European Southern Observatory, Alonso de C\'{o}rdova 3107, Casilla 19001, Vitacura, Santiago, Chile} 
\altaffiltext{9}{Department of Astronomy, Yale
  University, New Haven, CT, 06520-8101} 
\begin{abstract}
We present a catalog covering 1.62 deg$^2$ of the COSMOS/UltraVISTA field with PSF-matched photometry in 30 photometric bands.  The catalog covers the wavelength range 0.15$\micron$ $\rightarrow$ 24$\micron$ including the available $GALEX$, Subaru, CFHT, VISTA and $Spitzer$ data.  Catalog sources have been selected from the DR1 UltraVISTA K$_{s}$ band imaging that reaches a depth of K$_{s,tot}$ $=$ 23.4 AB (90\% completeness).  The PSF-matched catalog is generated using position-dependent PSFs ensuring accurate colors across the entire field.  Also included is a catalog of photometric redshifts ($z_{phot}$) for all galaxies computed with the EAZY code.  Comparison with spectroscopy from the zCOSMOS 10k bright sample shows that up to $z \sim$ 1.5 the $z_{phot}$ are accurate to $\Delta$$z$/(1 + $z$) = 0.013, with a catastrophic outlier fraction of only 1.6\%.  The $z_{phot}$ also show good agreement with the $z_{phot}$ from the NEWFIRM Medium Band Survey (NMBS) out to $z \sim$ 3.  A catalog of stellar masses and stellar population parameters for galaxies determined using the FAST spectral energy distribution fitting code is provided for all galaxies.   Also included are rest-frame U-V and V-J colors, L$_{2800}$ and L$_{IR}$.  The UVJ color-color diagram confirms that the galaxy bi-modality is well-established out to $z \sim$ 2.  Star-forming galaxies also obey a star forming ``main sequence" out to $z \sim$ 2.5, and this sequence evolves in a manner consistent with previous measurements.  The COSMOS/UltraVISTA K$_{s}$-selected catalog covers a unique parameter space in both depth, area, and multi-wavelength coverage and promises to be a useful tool for studying the growth of the galaxy population out to $z \sim$ 3 - 4.  
\end{abstract}
\section{Introduction}
The last decade has seen significant progress in our understanding of the evolution of massive galaxies (Log(M$_{star}$/$M_{\odot}$) $>$ 11.0) at 0 $< z <$ 5.  These galaxies exist in significant numbers already at $2 < z <$ 5, and their number density evolves fairly quickly over this redshift range \citep[e.g.,][]{Mclure2006,Marchesini2009, Marchesini2010, Cirasuolo2010, Caputi2011}.  This early and rapid formation of massive galaxies is not well-reproduced in current theoretical models, where high-mass galaxies typically form much later \citep[e.g.,][]{Fontanot2009,Guo2011,Henriques2012,Bower2012}.  
\newline\indent
Between 0 $< z <$ 2 the growth of massive galaxies is more gradual \citep[e.g.,][]{Bundy2006,Arnouts2007,Marchesini2009, Ilbert2010, Brammer2011, Bielby2012}, perhaps because this redshift range is also marked by a significant decrease in the star formation rates of these galaxies.  Current data show that the bi-modality between quiescent galaxies and star forming galaxies is largely established between 1 $ < z < $ 2 \citep[e.g.,][]{Williams2009,Whitaker2011,Brammer2011}.  Below $z <$ 1, it appears the transformation of massive star forming galaxies into massive quiescent galaxies is largely completed and any subsequent mass growth in the massive population is primarily driven by mergers \citep[e.g.,][]{Bundy2006,Arnouts2007, Ilbert2010, Brammer2011}.
\newline\indent
This impressive recent progress in measuring the evolution of massive galaxies has primarily been made possible by the extensive investments made in ground-based near infrared (NIR) imaging capabilities over the last decade.  The early deep pencil-beam NIR surveys that allowed for the first identifications of massive, high-redshift galaxies \citep[e.g.,][]{Franx2003,ForsterSchreiber2004,Daddi2005,vanDokkum2006} such as FIRES \citep[0.002 deg$^2$,][]{Labbe2003}, GOODS \citep[0.04 deg$^2$,][]{Wuyts2008} and MUSYC \citep[0.12 deg$^2$]{Quadri2007}, have now been superceded by much wider surveys that are equivalently deep, or deeper.  These surveys, such as the NMBS \citep[0.44 deg$^2$,][]{vanDokkum2009b,Whitaker2011}, the UKIDSS-UDS \citep[0.7 deg$^2$,][]{Lawrence2007,Williams2009}, and WIRDS \citep[2.1 deg$^2$,][]{McCracken2010,Bielby2012} are now the best-studied cosmic windows onto the massive galaxy population and have provided the source data for most of the results previously mentioned.
\newline\indent
The latest in this series of wider and deeper NIR sky surveys is the UltraVISTA survey \citep{McCracken2012}.  UltraVISTA has imaging in four broadband NIR filters (YJHK$_{s}$) as well as one narrow band filter centered on H$\alpha$ at $z = 0.8$ (NB118).  UltraVISTA is the deepest of the VISTA public surveys, and when fully complete will cover an area of 1.8 deg$^2$ down to K$_{s}$ $\sim$ 24.0, with a deeper component (referred to as ``ultra-deep") covering 0.75 deg$^2$ down to K$_{s}$ $\sim$ 25.6 \citep[see][]{McCracken2012}.    The current first data release of UltraVISTA is based on approximately one season of observing time and is now publicly available \citep{McCracken2012}.  In those data the imaging reaches a 5$\sigma$ depth of K$_{s}$ $<$ 23.9 AB in a 2$^{\prime\prime}$ aperture.  The full-depth UltraVISTA dataset will be acquired over a period of 5 -- 7 years.
\newline\indent
In addition to an unique combination of area and depth, one of the main strengths of UltraVISTA is that the survey field is the COSMOS field \citep{Scoville2007}.  COSMOS has arguably the most impressive array of multiwavelength coverage of any degree-scale part of the sky.  It contains X-ray data from $XMM-Newton$ \citep{Hasinger2007}, and $Chandra$ \citep{Elvis2009}, UV imaging from the $GALEX$ satellite \citep{Martin2005}, extensive optical broad-band and optical medium-band imaging from the CFHT and Subaru telescopes \citep{Taniguchi2007,Capak2007}, mid-infrared data from $Spitzer$ \citep{Sanders2007,Frayer2009}, submillimeter data from $Herschel$ \citep{Oliver2012}, millimeter data from AzTEC and MAMBO \citep{Scott2008,Aretxaga2011}, as well as radio observations from the VLA \citep{Shinnerer2007,Shinnerer2010}.  The field also has spectroscopy for $\sim$ 25,000 galaxies from the zCOSMOS-bright \citep{Lilly2007,Lilly2009} and zCOSMOS-deep surveys \citep{Lilly2007}.  In addition to this, it also has Hubble ACS data \citep{Koekemoer2007} covering the full field, as well as deep WFC3 NIR imaging and grism spectroscopy in part of the field from the CANDELS and 3D-HST surveys \citep{Koekemoer2011,Grogin2011,Brammer2012b}.  This extensive multi-wavelength coverage makes COSMOS an attractive field for performing studies of distant galaxies. 
\newline\indent
In this paper we present and make available a 30-band photometric catalog of the COSMOS field covering the wavelength range 0.15$\micron$ $\rightarrow$ 24$\micron$.  The selection of sources has been made using the high image-quality UltraVISTA K$_{s}$-band which allows for efficient selection of mass-complete samples of galaxies up to $z =$ 4.  In addition to the photometric catalog, we also make available a catalog of photometric redshifts, stellar population parameters, rest-frame colors, and UV/IR luminosities for all galaxies in the survey.  The stellar mass function of galaxies to $z =$ 4 determined using this catalog is presented in a companion paper \citep{Muzzin2013c}.
\newline\indent
The layout of this paper is as follows.  In $\S$ 2 we discuss the various observational datasets included in the photometric catalog.  In $\S$ 3 we outline the steps of the catalog creation such as the PSF matching, source detection, and data quality controls.  In $\S$ 4 we present the catalog of photometric redshifts and compare these to other photometric and spectroscopic redshift measurements.  In $\S$ 5 we present a catalog of stellar population parameters, and examine the basic properties of galaxies in the catalog such as the galaxy bi-modality and star formation ``main sequence".  We conclude with a summary in $\S$ 6.  Throughout this paper we assume an $\Omega_{\Lambda}$ = 0.7, $\Omega_{m}$ = 0.3, and H$_{0}$ = 70 km s$^{-1}$ Mpc$^{-1}$ cosmology.  All magnitudes are in the AB system. 
\begin{deluxetable}{lccl}
\tabletypesize{\scriptsize}
\scriptsize
\tablecaption{Summary of Photometric Data}
\tablehead{\colhead{Filter} & \colhead{ FWHM Seeing ($^{\prime\prime}$)} & \colhead{ 5$\sigma$ Depth (2.1$^{\prime\prime}$) } &  \colhead{ Reference }
\\
\colhead{(1)}& \colhead{(2)}& \colhead{(3)}& \colhead{(4)}
}
\startdata
FUV & 4.35. & 25.2 & Martin et al. (2005) \nl
NUV & 4.65. & 25.1 & -- \nl
  u$^{*}$ & 0.82 -- 0.89  &    26.6 -- 26.9   &   Capak et al. (2007) \nl
  B$_{j}$ & 0.71 -- 0.78  &    26.8 -- 27.1   &     -- \nl
 g$^{+}$  & 1.01 -- 1.20  &    26.3 -- 26.5   &     -- \nl
  V$_{j}$ & 0.74 -- 0.84  &    26.2 -- 26.4   &     -- \nl
 r$^{+}$  & 0.78 -- 0.85  &    26.2 -- 26.4   &     -- \nl
 i$^{+}$  & 0.53 -- 0.68  &    25.9 -- 26.1  &     -- \nl
 z$^{+}$  & 0.81 -- 0.91  &    25.0 -- 25.3   &     -- \nl
 IA427    & 0.66 -- 0.75  &    26.1 -- 26.2   &     -- \nl
 IA464    & 0.78 -- 1.05  &    25.8 -- 25.9   &     -- \nl
 IA484    & 0.54 -- 0.72  &    26.1 -- 26.2   &     -- \nl
 IA505    & 0.71 -- 0.84  &    25.8 -- 26.0   &     -- \nl
 IA527    & 0.59 -- 0.67  &    26.1 -- 26.2   &     -- \nl
 IA574    & 0.87 -- 0.94  &    25.7 -- 25.8   &     -- \nl
 IA624    & 0.70 -- 0.81  &    25.8 -- 26.0   &     -- \nl
 IA679    & 0.87 -- 1.02  &    25.5 -- 25.7   &     -- \nl
 IA709    & 0.76 -- 0.90  &    25.7 -- 25.9   &     -- \nl
 IA738    & 0.72 -- 0.80  &    25.6 -- 25.7   &     -- \nl
 IA767    & 0.98 -- 1.07  &    25.3 -- 25.5   &     -- \nl
 IA827    & 0.90 -- 1.08  &    25.4 -- 25.5  &     -- \nl
  Y       & 0.82 -- 0.86  &    24.3 -- 24.6  &    McCracken et al. (2012) \nl
  J       & 0.81 -- 0.85  &    24.2 -- 24.4  &     -- \nl
  H       & 0.78 -- 0.82  &    23.8 -- 24.1   &     -- \nl
 K$_{s}$  & 0.77 -- 0.82  &    23.7 -- 23.9  &     -- \nl
3.6$\micron$  & 1.75  &    23.9   &    Sanders et al. (2007) \nl
4.5$\micron$  & 1.78  &    23.3   &    -- \nl
5.8$\micron$  & 1.99  &    21.3   &    -- \nl
8.0$\micron$  & 2.24  &    21.0   &    -- \nl
24$\micron$   & 5.91  &    45 $\mu$Jy   &    -- \nl
\enddata
\tablecomments{Seeing information is the full range of seeing from an average of 10 PSF stars in each of the 9 subfields.  The depths within the 2.1$^{\prime\prime}$ aperture are on the PSF-matched images and hence are effective depths for the purpose of color measurements in the catalog. The quoted depth at 24$\micron$ is a total flux.}
\end{deluxetable}
\begin{figure}
\plotone{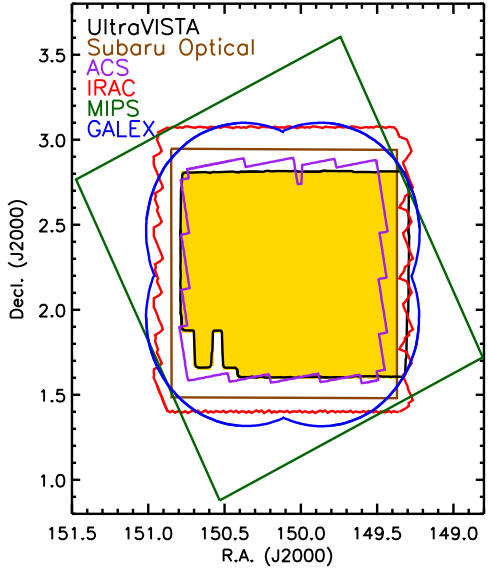}
\caption{\footnotesize Coverage diagram for the UltraVISTA/COSMOS field.  The filled gold region shows the area covered by the K$_{s}$-selected catalog.  The catalog contains objects only in regions where both NIR and optical coverage is available.  It covers a total of 1.62 deg$^{2}$, when regions contaminated by bright stars are excluded.}
\end{figure}
\section{Description of the Dataset}
\begin{figure*}
\plotone{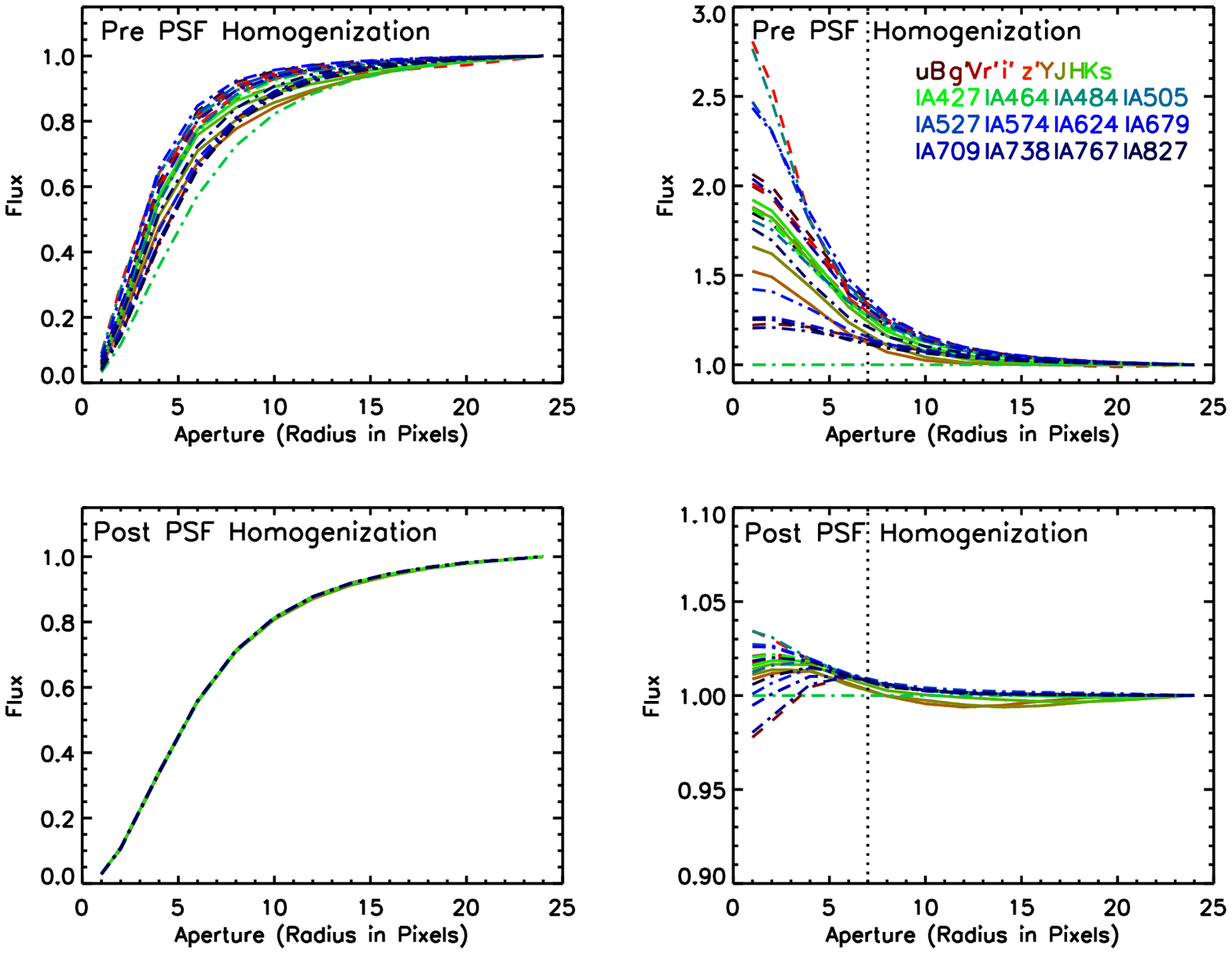}
\caption{\footnotesize Top left: Curve of growth for bright PSF stars in the field.  Subaru optical broad bands (dashed lines), NIR bands (solid lines) and Subaru optical medium bands (dot-dashed lines) are shown.  Top right: Same curves of growth normalized relative to the worst seeing image (IA464, $\sim$ 1.05$^{\prime\prime}$).  The color aperture is shown as the dotted vertical line.  Bottom panels: Same as top panels but after PSF homogenization has been performed. Before PSF homogenization the range of fluxes within the color aperture is a factor of 1.4.  After PSF homogenization colors within the color aperture are accurate to $\sim$ 1\%.}
\end{figure*}
\subsection{Photometric Datasets}
There is a wealth of imaging data at various wavelengths available for the COSMOS field.  For the K$_{s}$-selected catalog we have chosen to include 30 photometric bands that cover the wavelength range 0.15$\micron$ $\rightarrow$ 24$\micron$.  These bands and the papers that describe these datasets are summarized in Table 1.  
\newline\indent
The catalog is based on the YJHK$_{s}$ NIR imaging data from UltraVISTA \citep{McCracken2012}, and the inclusion or exclusion of the available datasets was chosen to provide a match in areal coverage and depth of those data.  The optical data consists of broadband data taken with Subaru/SuprimeCam (g$^{+}$r$^{+}$i$^{+}$z$^{+}$B$_{j}$V$_{j}$), as well as u$^{*}$ data from the CFHT/MegaCam \citep[][]{Taniguchi2007,Capak2007}.  We have also included the 12 optical medium bands (IA427 -- IA827) from Subaru/SuprimeCam \citep[][]{Capak2007} for a total of 23 optical/NIR bands.  Observations from the $GALEX$ FUV and NUV channels \citep[see][]{Martin2005}, as well as the 3.6$\micron$, 4.5$\micron$, 5.8$\micron$, 8.0$\micron$ and 24$\micron$ channels from $Spitzer's$ IRAC+MIPS cameras \citep[see][]{Sanders2007} have also been included using a source-fitting technique designed for determining robust colors in highly-blended imaging data (see $\S$ 3.5).
\newline\indent
Several datasets have not been incorporated into the catalog.  These include the deep u$^{*}$g$^{\prime}$r$^{\prime}$i$^{\prime}$z$^{\prime}$ imaging taken as part of the CFHTLS-Deep survey.  These data are deeper than the Subaru broadband data; however, they cover $\sim$ 0.9 deg$^2$ near the center of the COSMOS field, which is only slightly more than half of the UltraVISTA area.  In principle, these data should improve the quality of the photometric redshifts and stellar mass measurements for galaxies in that region of the survey; however, including them would cause the uncertainties in those quantities to be location-dependent.  This makes it much more difficult to estimate uncertainties in quantities derived from the entire survey such as the stellar mass function or the color distribution of galaxies.  For the same reason we have chosen to not include the deep NIR medium band imaging from the NMBS \citep{Whitaker2011}, which covers only 0.22 deg$^2$ of the field.  
\newline\indent
We have also not included the J-band imaging of COSMOS from KPNO \citep{Capak2007}, nor the H and K$_{s}$ imaging from CFHT/WIRCAM \citep{McCracken2010}.  Those data have full coverage of the field; however, they use similar filters as those in UltraVISTA and are shallower.
\subsection{Field Geometry}
\indent
The geometry of the COSMOS field is roughly a square patch on the sky; however, the coverage from the various datasets does not overlap perfectly.  In Figure 1 we plot a schematic view of the layout of the datasets used in the photometric catalog.  The selection band for the catalog is the UltraVISTA K$_{s}$-band which is shown as the black outline in Figure 1.  The UltraVISTA full field covers $\sim$ 1.8 deg$^2$; however, as Figure 1 shows it is placed slightly to the west of the centroid of the Subaru broad and medium band imaging (orange outline).  This offset in the UltraVISTA field was necessary to ensure that the four ``strips" that are observed as part of the ultra-deep component of UltraVISTA did not coincide with the positions of bright stars.  The effective area of overlap between the Subaru optical data and the UltraVISTA NIR data is 1.62 deg$^2$ once bright stars have been masked, and this region is shown as the gold shaded region in Figure 1.  All sources in the catalog are contained within this overlap region. 
\newline\indent
As Figure 1 shows, there is complete coverage for the 1.62 deg$^2$ catalog region from $GALEX$, IRAC, and MIPS.  The majority of the ACS coverage is also within the UltraVISTA area. 
\section{Photometry and Source Detection}
\indent
Measuring the colors of galaxies accurately is paramount for determining properties such as photometric redshifts and stellar masses.  The available imaging data in the COSMOS/UltraVISTA field comes with a wide range of Point Spread Function (PSF) shapes and sizes and these variations need to be accounted for in the color measurements.  The range of image qualities for the various filters is listed in Table 1 \citep[see also][]{Capak2007}.   As Table 1 shows, the image quality ranges from as good as 0.5$^{\prime\prime}$ full width half maximum (FWHM) in some of the optical bands, to as poor as 4$^{\prime\prime}$ -- 5$^{\prime\prime}$ FWHM in the $GALEX$ and MIPS 24$\micron$ imaging.    The optical and NIR imaging have PSFs that are comparable in shape (i.e., roughly Gaussian) and FWHM (0.5$^{\prime\prime}$ - 1.2$^{\prime\prime}$) and therefore PSF matching of those bands is performed in a similar way using standard techniques (see $\S$ 3.1).  The space-based imaging from $GALEX$, IRAC and MIPS have more complicated PSF shapes that have considerable wings, as well as a much larger FWHM.  Photometry for those bands is performed separately using a source-fitting code designed to measure accurate photometry for highly-blended sources and is described in $\S$ 3.5.  
\subsection{Optical and NIR PSF Matching}
PSF matching between the optical and NIR bands is performed by degrading the image quality of all bands to the image quality of the worst-seeing band.  This process is performed separately in different regions of the survey. Region-dependent PSF matching is needed because full coverage of the COSMOS field required 9 SuprimeCam pointings per filter \citep[see][]{Capak2007}.  The result of these multiple pointings is that in addition to the image-quality variations between filters, there are also image-quality variations from pointing-to-pointing within a given filter.  The range of PSF FWHMs for each filter is listed in Table 1.  
\newline\indent
To perform the PSF matching we divided the survey into 9 patches (labeled COSMOS-1 -- COSMOS-9) closely tied to the positions of the SuprimeCam pointings.  We note that this approach does not account for intra-stack seeing variations in the NIR bands which are caused by the sparsely-filled nature of VIRCAM mosaic.  The intra-stack PSF variations of the order a few hundredths of an arcsecond \citep[see e.g.,][]{McCracken2012}, and hence are much smaller than the PSF differences between different bands.  
\newline\indent
Within each region 10 bright, unsaturated reference PSF stars were chosen and coadded into a reference PSF for that filter/region.  Stars have a range of colors, and therefore it was difficult to find PSF stars that were suitably bright but unsaturated from the $u^{*}$ $\rightarrow$ K$_{s}$ bands.  This was dealt with by choosing one set of stars for the $u^{*}$ $\rightarrow$ $IA679$ bands, and a separate set of stars for the $IA738$ $\rightarrow$ K$_{s}$ bands.  An additional set of PSF stars was chosen for the Subaru i$^{+}$ band.  The i$^{+}$ imaging has superior image quality ($\sim$ 0.5$^{\prime\prime}$) across the field, and was taken with longer exposure times than the other bands so all stars down to i$^{+}$ $<$ 21.8 are saturated \citep[see also][]{Capak2007}.  Finding unsaturated PSF stars required going to much fainter stars than in the other bands.
\newline\indent
With a reference PSF constructed for each band/region we used the task \texttt{lucy} from the IRAF STSDAS package to compute the convolution kernels necessary to degrade the images to the image quality of the worst-seeing band.  In 2/9 of the regions the worst image-quality band was the IA464 band, with typical seeing of $\sim$ 1.05$^{\prime\prime}$.  In the other 7/9 regions the worst image-quality band was the g$^{+}$ band with seeing of 1.0$^{\prime\prime}$ -- 1.2$^{\prime\prime}$.
\newline\indent
In Figure 2 we plot a illustration of the PSF matching process for the COSMOS-1 field.  In the upper left panel we plot the growth curve of the PSF stars before PSF homogenization is performed.  In the upper right panel we plot the same growth curves normalized relative to the worst-seeing band (IA464).  The dotted vertical line shows the 2.1$^{\prime\prime}$ aperture used for color measurements.  Prior to PSF homogenization, the dispersion in the flux contained within the color aperture for the best and worst image-quality bands is a factor of $\sim$ 1.4.
\newline\indent
In the bottom panels of Figure 2 we plot the reference stars after PSF homogenization has been performed.  As the right panel shows, the dispersion in flux within the color aperture for the reference PSFs is reduced to $<$ 1\%.  
\begin{figure*}
\plotone{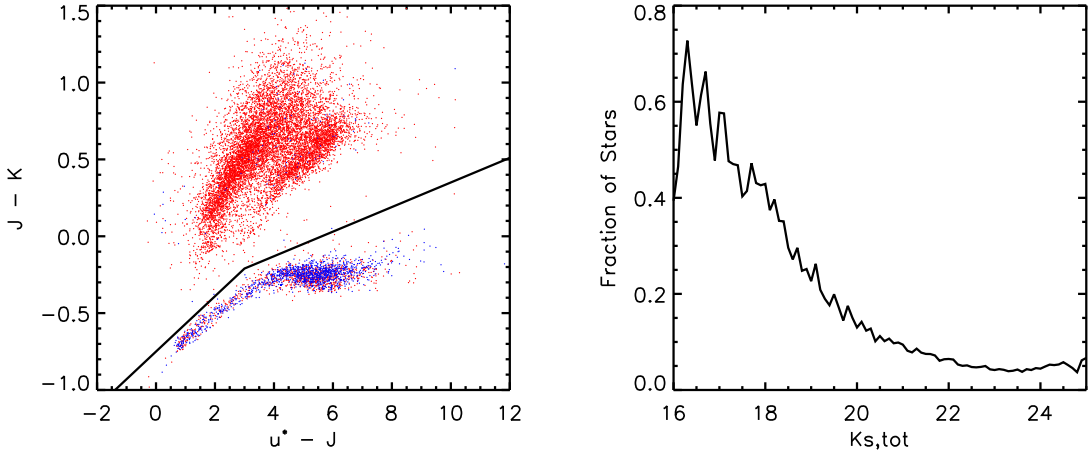}
\caption{\footnotesize Left panel: The color -- color space used to perform star/galaxy separation.  One-third of objects with K$_{s}$ $<$ 21.5 are plotted.  Objects that are classified as stars using SExtractor (\texttt{class\_star} $>$ 0.95) are plotted in blue and objects classified as galaxies (class\_star $<$ 0.95) are plotted in red.  There is a clear separation of stars and galaxies in the u$^{*}$ - J vs. J - K$_{s}$ color -- color space.  In the catalog, objects that lie below the lines are classified as stars, and those above are classified as galaxies.  Right panel: Fraction of objects that are stars as a function of K$_{s}$-band magnitude. }
\end{figure*}  
\subsection{Source Extraction and Photometry}
Source detection and photometry is performed using the SExtractor package \citep{Bertin1996} in dual image mode with the non-PSF-matched UltraVISTA K$_{s}$-band image as the detection image.  Objects are detected by flagging pixels $>$ 1.7$\sigma$ above the background, after a 2 $\times$ 2 pixel convolution kernel has been applied.  We require 10 connected pixels for a detection.  At face-value this appears to be a conservative detection limit; however, it is required to prevent the detection of a significant number of spurious sources.  The native pixel scale of VIRCAM is 0.34$^{\prime\prime}$ pixel$^{-1}$ but the data have been resampled to the COSMOS optical data pixel scale of 0.15$^{\prime\prime}$ pixel$^{-1}$ \citep[see][]{McCracken2012}.  This makes the noise in the UltraVISTA images highly correlated causing SExtractor to underestimate it and return a high fraction of spurious detections if a more relaxed object detection criterion is used.  
\newline\indent
Photometry is performed in each of the 9 regions independently and afterwards these are merged into a master catalog for the entire field.  For each filter we determine two fluxes, a Kron-aperture flux based on SExtractor's \texttt{flux\_auto} parameter, as well as the flux within a circular color aperture.  The color aperture was chosen to be 14 pixels in diameter, corresponding to 2.1$^{\prime\prime}$ on the sky.  This size makes the color aperture $\sim$ 2$\times$ the FWHM of the worst-seeing image in each region.  
\newline\indent
For each galaxy we supply a total K$_{s}$-band magnitude (K$_{s,tot}$) based on the \texttt{flux\_auto} parameter from SExtractor.  This flux is the flux measured within 2.5 times the Kron radius \citep[R$_{K}$,][]{Kron1980}.  The measured flux within 2.5 $\times$ R$_{K}$ should account for $>$ 96\% of the total flux of the galaxy \citep{Kron1980}.  We correct the \texttt{flux\_auto} to a total flux by measuring the growth curve of bright stars out to a radius of 8$^{\prime\prime}$.  Depending on the R$_{K}$, this correction is of the order 2 - 4\%, as expected.
\newline\indent
In the catalog we remove all objects with magnitudes $<$ 3$\sigma$ limit in the K$_{s}$-band (K$_{s}$ = 24.35, 2.1$^{\prime\prime}$ aperture), which results in a final photometric catalog of 262 615 sources.
\subsubsection{Galactic Extinction Correction}
The photometry in all bands is corrected for Galactic dust attenuation using the dust maps from \cite{Schlegel1998}.  Dust extinctions are calculated in each of the 9 photometry regions individually.  The COSMOS field does not have substantial Galactic dust, and the corrections are of order 15\% in the $GALEX$ bands, 5\% in the optical and $<$ 1\% in the NIR and MIR.
\subsubsection{Photometric Errors}
All of the optical and NIR images have been resampled during the mosaicking process.  The resampling causes correlations in the noise between pixels and therefore standard Poission estimation of the background noise is not a reliable estimate of the photometric errors.  We measure the background noise in each band, and in each region, by placing 10 000 empty apertures across the field and measuring the rms flux in those empty apertures.  These rms estimates also give a measure of the 5$\sigma$ depth for each filter, and we list those depths and the range of depths in Table 1.  The quoted depths in Table 1 are the 5$\sigma$ depth in the 2.1$^{\prime\prime}$ color aperture, and hence are the effective depth of the photometry.  Given the uncertainty in the absolute IRAC zeropoint, we have also included an additional 5\% systematic error to the IRAC magnitudes in quadrature.
\newline\indent
Errors in K$_{s,tot}$ are calculated using the method developed by \cite{Labbe2003}, which as also used by \cite{Quadri2007} and \cite{Whitaker2011}.  The background noise in an aperture of diameter N pixels will scale as N$^{2}$ in the limiting case of perfect correlation between pixels, and will scale as N in the limiting case of no correlation between pixels.  We parameterize the background noise in an aperture as $\sigma_{ap}$ = $\sigma_{1}\alpha$N$^{\beta}$, where $\sigma_{1}$ is the size of the aperture in arcseconds, and 1 $< \beta <$ 2.  We perform the empty-aperture measurement for a range of aperture sizes and fit this relation for $\alpha$ and $\beta$, finding $\beta$ = 1.82 and $\alpha$ = -0.35.  From this relation we then compute the error in the \texttt{flux\_auto} based on the rms noise in an aperture of 2.5R$_{K}$.
\subsection{Star Galaxy Separation}
\indent
Similar to the NMBS, star galaxy separation is performed use the J - K$_{s}$ vs. u$^{*}$ - J color space.  In Figure 3 we plot this color space  for a randomly-selected subsample of 30\% of the objects with K$_{s}$ $<$ 21.5.  Points are color-coded by SExtractor's \texttt{class\_star} parameter, with star-like profiles in blue, and galaxy-like profiles in red.  As Figure 3 shows, the two types of objects clearly segregate in color space.  Based on this segregation, objects are classified as galaxies if they meet the following color criteria,
\begin{equation}
J - K_{s} > 0.18 \times (u^* - J) - 0.75, [u - J < 3.0],
\end{equation}
\begin{equation}
J - K_{s} > 0.08 \times (u^* - J) - 0.45, [u - J > 3.0]
\end{equation}
\indent
As shown in the right panel of Figure 3, stars dominate the catalog at K$_{s,tot}$ $<$ 17, but make up $<$ 10\% of objects at K$_{s,tot}$ $>$ 21.0.
\subsection{Survey Completeness}
\indent
In order to construct mass- or luminosity-limited samples, the completeness as a function of the selection band needs to be quantified.  To estimate the point-source completeness we use the PSF stars as template sources, and insert these into the original non-PSF-matched K$_{s}$-band image.  We then attempt to recover these sources with SExtractor.  When recovering the simulated stars we used the identical SExtractor parameters as were used for object detection in the catalog.  
\newline\indent
This completeness test is performed in two ways.  Firstly, we insert sources into a version of the image where all objects have been masked out using SExtractor's segmentation map, and then attempt to recover these objects with SExtractor.  The recovery rate of these objects gives an estimate of the completeness based solely on the noise properties of the image.  We then perform the test again, this time using the real image with all objects still included.  The recovery rate of this method gives a measure of the overall completeness, which is always less than unity because of real effects that cause objects to be missed such as contamination from bright stars, or blending with other galaxies.  
\newline\indent
In Figure 4 we plot the completeness curves as a function of K$_{s,tot}$.  The 90\% completeness limit in terms of the noise properties alone is K$_{s,tot}$ $=$ 24.0 mag; however, this corresponds to $<$ 80\% completeness in the real data.  Based on the noise properties, the 100\% completeness level of the UltraVISTA K$_{s,tot}$ data is 23.4 mag.  This K$_{s,tot}$ limit corresponds to a 90\% completeness level in the actual data.  At magnitudes fainter than 23.4 the data becomes incomplete quite rapidly.  This can be seen as the inflection point in the number counts in the inset of Figure 4 \citep[see also][]{McCracken2012}, which is the result of the increasing incompleteness as well as the increase in photometric uncertainties due to poor S/N near the completeness limit.  In order to construct mass-complete samples we only use sources in the catalog with K$_{s,tot}$ $\leq$ 23.4.  We note that this completeness is 0.08 mag {\it brighter} than would be measured using the DR1 UltraVISTA data from \cite{McCracken2012}.  We have adjusted the zeropoint of the DR1 images by this amount in order to bring it into better agreement with other K$_{s}$ surveys such as 2MASS and the NMBS (see Appendix).  
\newline\indent
Also shown in the inset of Figure 4 is the sample completeness measurement from the NMBS \citep{Whitaker2011}.  The 90\% completeness of the NMBS is K$_{tot}$ $<$ 22.8, showing that UltraVISTA is $\sim$ 0.6 mag deeper than NMBS.  The surveys have similar exposure times on 4m-class telescopes, so the extra depth is primarily a result of the superior image quality in the UltraVISTA DR1 \citep[$\sim$ 0.75$^{\prime\prime}$,][]{McCracken2012} compared to the NMBS \citep[$\sim$ 1.1$^{\prime\prime}$,][]{Whitaker2011}.  
\begin{figure}
\plotone{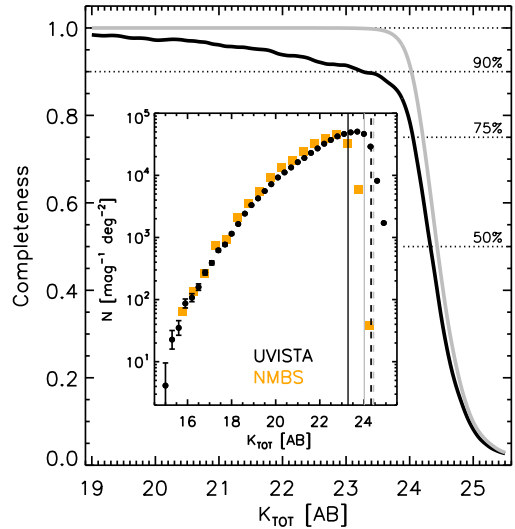}
\caption{\footnotesize Completeness fraction as a function of total K$_{s}$-band flux for point sources.  The completeness is calculated using simulations which insert point sources into the K$_{s}$-band image and attempt to recover them.  The grey curve is calculated by masking all real sources in the image and demonstrates the completeness based on the background noise in the image.  The black curve is calculated by keeping in all real sources and therefore is lower because of effects such as blending and confusion.  In the inset we show the differential number counts in K$_{s}$-band and a comparison to the number counts from the NMBS \citep{Whitaker2011}.  Based on the simulations and the inflection in the number counts, the 90\% completeness limit of the catalog is K$_{s,tot}$ $<$ 23.4.}
\end{figure}
\subsection{IRAC and MIPS Photometry}
\indent
The PSF FWHM of IRAC and MIPS are approximately 2 -- 5$\times$ larger than the FWHM of the worst-seeing optical and NIR PSF FWHMs.  This means that degrading the image quality of the optical/NIR data to that image quality would cause a substantial reduction in the signal-to-noise (S/N) of the color measurements as well as unnecessary blending of well-resolved galaxies.  Therefore, instead of degrading PSFs we use a source-fitting code developed to measure deblended photometry of heavily-confused images.  This code is well-tested and has been used in many previous K$_{s}$-selected catalogs \citep[see][]{Labbe2005,Wuyts2007,Marchesini2009,Williams2009,Whitaker2011,Marchesini2012}, including several with ultra-deep IRAC imaging \citep{Labbe2010,Labbe2012}.
\newline\indent
In brief, we assume that there are no color gradients in galaxies between the K$_{s}$ band and the IRAC and MIPS bands.  The K$_{s}$ band is then used as a high-resolution template image to deblend the IRAC and MIPS photometry.  Each source extracted from the K$_{s}$ image is convolved with a kernel derived from bright PSF stars in the K$_{s}$ and IRAC/MIPS images.  The convolved galaxies are then fit as templates in the IRAC and MIPS bands with the total flux left as a free parameter.  In this process, all objects in the image are fit simultaneously.  Once the template fitting is converged, a ``cleaned" image is produced for each object in the catalog by subtracting off all nearby sources \citep[for an example of this process see][Figure 1]{Wuyts2007}.  Aperture photometry is then performed on the cleaned image of each source.  For the IRAC (MIPS) channels the photometry is performed in a 3$^{\prime\prime}$ (5$^{\prime\prime}$) diameter aperture for each object.  This flux is then corrected to a flux within the 2.1$^{\prime\prime}$ color aperture via,
\begin{equation}
F_{IRAC}(2.1^{\prime\prime}) = F_{IRAC,cleaned}(3^{\prime\prime}) \times \frac{K_{conv,worst-PSF}(2.1^{\prime\prime})}{K_{conv,IRAC}(3^{\prime\prime})}, 
\end{equation}
where K$_{conv,worst-PSF}(2.1^{\prime\prime})$ is the flux in the PSF-matched K$_{s}$ image in the color aperture and K$_{conv,IRAC}(3^{\prime\prime})$ is the flux in the K$_{s}$ image convolved to the IRAC PSF.  Given that MIPS fluxes are primarily used to indicator SFRs, and are not used for colors, within the catalog the MIPS fluxes are listed as total fluxes rather than color fluxes.  The MIPS fluxes have been converted to total fluxes using an aperture correction of a factor of 3.7, as listed in the MIPS instrument handbook\footnote{http://irsa.ipac.caltech.edu/data/SPITZER/
docs/mips/mipsinstrumenthandbook}.
\subsection{GALEX Photometry}
\indent
The image quality of the $GALEX$ NUV and FUV data are also substantially poorer than that available from the ground-based data ($\sim$ 4$^{\prime\prime}$ - 5$^{\prime\prime}$).  We cleaned and photometered the $GALEX$ data using the same source-fitting code as for the IRAC and MIPS data.  For the $GALEX$ data we use the CFHT u$^{*}$-band data as a template rather than the K$_{s}$-band because the u$^{*}$ band and the FUV and NUV bands have a close correspondence in wavelength.  This required creating a new source list based on the u$^{*}$ images and then matching that source list to the K$_{s}$ source list.  When performing this matching we use a conservative matching tolerance ($<$ 0.5$^{\prime\prime}$) to ensure only true counterparts are matched to the K$_{s}$-selected catalog.  No $GALEX$ photometry is provided for K$_{s}$-selected sources that do not have a clear match to the u$^{*}$-selected sources.  We note that only the $GALEX$ photometry is provided by matching sources, the u$^{*}$-band photometry itself is still measured using the standard PSF matching method and SExtractor in dual-image mode.
\begin{deluxetable*}{lll}
\tabletypesize{\footnotesize}
\scriptsize
\tablecaption{Summary of Photometric Catalog}
\tablehead{\colhead{Column} & \colhead{ Parameter Name } & \colhead{ Description } 
}
\startdata
1 & id & Object identifier number \nl
2, 3 & ra, dec & Right Ascension and Declination in J2000 decimal degrees \nl
3, 4 & xpix, ypix & Pixel position of object in the K$_{s}$ image \nl
5 & Ks\_tot & Total K$_{s}$-band flux with additional aperture correction applied \nl
6 & eKs\_tot & Error in total K$_{s}$-band magnitude determined from scaled empty apertures \nl
7 -- 67 & X, eX & Flux and error in filter X measured in a 2.1$^{\prime\prime}$ aperture from PSF-matched images \nl
68 & K\_flag & SExtractor's FLAG output for the K$_{s}$-band image \nl
69 & K\_star & SExtractor's CLASS\_STAR output from the K$_{s}$ band image \nl
70 & K\_Kron & Kron radius in the K$_{s}$-band \nl
71 & apcor & Aperture correction that has been applied to FLUX\_AUTO to determine Ks\_tot \nl
72, 73, 74 & z\_spec, z\_spec\_cc, z\_spec\_id & Spectroscopic redshift, spectroscopic redshift quality flag, and ID number from the zCOSMOS catalog \nl
75 & star & Star/galaxy indicator determined from color-color plot (star = 1, galaxy = 0) \nl
76 & contamination & Indicates proximity to a bright star in the optical or NIR (contaminated = 1, uncontaminated = 0) \nl
77 & nan\_contam & Number of filters where object lies near gaps in photometry or near a saturated star \nl
78, 79 & orig\_cat\_id, orig\_cat\_field & ID and tile of object in the tile catalog \nl
80 & USE & Indicates galaxies with uncontaminated photometry and S/N $>$ 5 \nl
\enddata
\end{deluxetable*} 
\subsection{zCOSMOS Spectroscopic Redshifts}
\indent
We match the photometric catalog to the catalog of spectroscopic redshifts ($z_{spec}$) available from the zCOSMOS 10k-bright sample \citep{Lilly2009}.  That catalog contains $\sim$ 10 000 redshifts for galaxies with i$^{\prime}$ $<$ 22.5.  Due to the bright limit, the majority of redshifts in the catalog are at 0.1 $< z <$ 1.5.  The matching is performed for the subset of zCOSMOS sources with redshift confidence classes in the range 3 $<$ CC $<$ 5.  These are the highest-confidence spectroscopic redshifts, and should be 99\% accurate \citep{Lilly2007}.  We match only the high-confidence $z_{spec}$ because these are used to determine offsets in the zeropoints in various filters ($\S$ 4.1), and to determine quantities such as stellar masses and rest-frame colors.  Uncertain $z_{spec}$ can be extremely wrong resulting in unphysical results for this process. 
\newline\indent
Matching is done using an 0.5$^{\prime\prime}$ search radius.  The optical i$^{+}$-band and the UltraVISTA K$_{s}$-band are well-registered so unique matches are found for all sources.  Overall, there are 5105 $z_{spec}$ from zCOSMOS matched to sources in the K$_{s}$-selected catalog. 
\subsection{Corrections for Bright Stars and Bad Regions}
The COSMOS field is large enough that it contains regions with very bright stars.  These bright stars create reflections and large diffraction spikes that make the photometry for galaxies in those regions unreliable.  In the catalog we provide a parameter, \texttt{contamination} which indicates whether an object's photometry has been contaminated by a nearby bright star.  The contamination is determined by first generating a source list of optically-bright stars within the COSMOS field from the USNO-B catalog, as well as a list of NIR-bright stars from the 2MASS catalog \citep{Skrutskie2006}.  We determined an empirical relation between the brightness of the star and radius of the contaminated photometry.  All objects that are within the contamination radius of a bright star in either the optical or NIR are then given a contamination flag = 1.  
\newline\indent
In general we have been conservative with the contamination radius in order to provide the most reliable photometric catalogs possible.  Still, the total area considered to be contaminated by bright stars is fairly modest.  The full coverage region of the catalog covers an area of 1.68 deg$^2$, whereas the usable, uncontaminated region covers 1.62 deg$^2$.  This means only $\sim$ 4\% of the total area is lost to bright stars.
\newline\indent
The Subaru optical bands also contain some small square-like regions throughout the survey that do not have data.  Some of these regions are located at the central few pixels of bright stars, but others occur throughout the survey area, seemingly at random locations.  It is unknown what the source of these regions are, but they occur most frequently in the i$^{+}$ band data.  The pixel values in these regions have either been set to ``nan" or 10$^{-31}$, and failure to mask these regions causes nonsensical fluxes to be measured for nearby galaxies.  The regions occur frequently enough that we developed a procedure for identifying them.
\newline\indent
A pixel map of each optical band showing the location of the ``nan" region is made.  This map is then smoothed with a kernel that grows the size of the contaminated regions by a factor of $\sim$ 2.  When the final catalog is constructed these mask regions are then checked for each object, in each filter.  If the object lies within a contaminated region for that filter, the flux in that filter is set to -99, and the parameter \texttt{nan\_contam} is incremented by 1.  This setting for the flux makes the photometric redshift fitting code ($\S$ 4) and stellar population fitting code ($\S$ 5) effectively consider the object not-observed in that particular filter.  This is useful because the nan regions are in different locations in various filters.  The filter-by-filter masking allows us to effectively keep objects with some contamination in the catalog, but remove only the photometry in the contaminated filters.  Because \texttt{nan\_contam} is incremented, it is a metric of the number of filters that have contamination for a given object.  It is recommended to only use objects with contamination in $<$ 5 filters. 
\subsection{Photometric Catalog Layout and Useage}
\indent
The layout of the full photometric catalog is summarized in Table 2.  The catalog is presented as a set of fluxes in the 2.1$^{\prime\prime}$ color aperture with an AB zeropoint of 25.0.  These can be converted to AB magnitudes via m$_{x}$ = -2.5log$_{10}$(f$_{x}$) + 25.0, where x denotes a given filter.  Also given is the total K$_{s}$-band magnitude from SExtractor's \texttt{flux\_auto}, which has been corrected to a total flux using the growth curve of the PSF stars.  The flux in any filter can be converted to a total flux via,
\begin{equation}
f_{x,tot} = f_{x} \times \frac{f_{Ks,tot}}{f_{Ks}}.
\end{equation}
\indent
The catalog contains the \texttt{star} indicator (= 1 for stars, = 0 for galaxies), as well as the \texttt{contamination} and \texttt{nan\_contam} parameters described in $\S$3.8.  Lastly, we include a parameter \texttt{USE}.  Selecting objects with \texttt{USE} = 1 in the catalog selects the subset of objects with \texttt{star} = 0, \texttt{contamination} = 0, \texttt{nan\_contam} $<$ 5, and K$_{s}$ $<$ 23.9.  The latter is the 5$\sigma$ depth of the survey in the color aperture.  Objects with \texttt{USE} = 1 are considered to be galaxies with uncontaminated  photometry with fluxes sufficiently bright that the photometry is still accurate.
\section{Photometric Redshifts}
\indent
Photometric redshifts ($z_{phot}$) are calculated for all galaxies using the EAZY software \citep{Brammer2008}.  EAZY determines the $z_{phot}$ for galaxies by fitting their SEDs to linear combinations of seven templates, six of which are derived from the PEGASE models \citep{Fioc1999},  as well as an additional red template from the models of \cite{Maraston2005}. A detailed description of EAZY's fitting process can be found in \cite{Brammer2008}.  EAZY also accounts for potential mismatches between the data and the templates via the ``template error function" \citep[see][]{Brammer2008}.  The template error function weights photometric points in the template fitting based on their implied rest-frame wavelength.  In particular, when using the default template error function, measurements corresponding to the rest-frame NIR are down-weighted compared to the rest-frame optical given the current uncertainties in models at these wavelengths \citep[e.g.,][]{Maraston2005,Kriek2010}. 
\newline\indent
Photometric redshifts were determined with EAZY primarily using the default set of parameters, although several optimizations were added after examination of the initial output.   For the first run we used the 7 default templates, the v1.0 template error function, the K$_{s,tot}$ magnitude prior, and allowed photometric redshift solutions in the range 0 $< z <$ 6.  
\newline\indent
Comparison of the resulting $z_{photo}$ with the z$_{spec}$'s from zCOSMOS was good, although there were some significant outliers ($\sim$ 5\%).  Some experimentation showed that catastrophic outliers could largely be eliminated by running EAZY with a 5\% systematic error included for all photometric bands.  The 5\% systematic helps reduce catastrophic outliers because the optical medium bands have narrow bandpasses and therefore problems in the photometry in consecutive bands can create very sharp features in the SED.  
\newline\indent
After eye-examination of the best-fit models for a subsample of the galaxies it became apparent that there were two populations of galaxies that were not well-described with the default template set.  The first of these were galaxies at $z >$ 1 with post starburst-like SEDs.  These galaxies tend to have strong Balmer breaks, but also a continuum that is very blue redward of the Balmer break, and it is not possible to reproduce such an SED with the default EAZY template set.  We added a 1 Gyr old single-burst \cite{Bruzual2003} model to the template set and this significantly improved the fits for these galaxies.  The need for a similar template in EAZY was also reported by \cite{Onodera2012}.
\newline\indent
The second set of galaxies with problematic fits was a population of UV-bright galaxies at 1.5 $< z <$ 3.5 with UV continua that were blue, but still redder than could be produced with the default templates.  Given that the catalog is K$_{s}$-selected, these are likely to be the most massive part of the Lyman Break Galaxy (LBG) population, and they may have slightly more dust extinction than lower-mass LBGs.  We added a slightly dust-reddened young population to the template set and this improved the fit to this population.  
\begin{deluxetable}{lc}
\tabletypesize{\footnotesize}
\scriptsize
\tablecaption{Zeropoint Offsets Determined from Spectroscopic Redshifts}
\tablewidth{1.5in}
\tablehead{\colhead{Filter} & \colhead{ Offset } \nl
\colhead{} & \colhead{ (Mag) }
}
\startdata
fuv & -- \nl
nuv & -- \nl
  u$^{*}$ & -0.10  \nl
  B$_{j}$ & -0.12  \nl
 g$^{+}$  & -0.11 \nl
  V$_{j}$ & 0.03  \nl
 r$^{+}$  & -0.06  \nl
 i$^{+}$  & -0.03  \nl
 z$^{+}$  & -0.06  \nl
 IA427    & -0.12  \nl
 IA464    & -0.12 \nl
 IA484    & -0.08  \nl
 IA505    & -0.07  \nl
 IA527    & -0.09  \nl
 IA574    & -0.17  \nl
 IA624    & -0.09  \nl
 IA679    & 0.06  \nl
 IA709    & -0.10  \nl
 IA738    & -0.15  \nl
 IA767    & -0.13  \nl
 IA827    & -0.17  \nl
  Y       & -0.10  \nl
  J       & -0.14  \nl
  H       & -0.18  \nl
 K$_{s}$  & -0.08\tablenote{The K$_{s}$ zeropoint offset is determined by comparing photometry with other surveys, not from the spectroscopic redshifts}   \nl
3.6$\micron$  & --  \nl
4.5$\micron$  & --  \nl
5.8$\micron$  & --  \nl
8.0$\micron$  & -- \nl
24$\micron$   & --  \nl
\enddata
\tablecomments{Offsets are defined such that ZP$_{EAZY}$ = ZP$_{nominal}$ + offset.}
\end{deluxetable}
\subsection{Zeropoint Offsets}
\indent
Photometric redshifts are extremely sensitive to errors in photometric zeropoints.  This is especially pronounced when medium bandwidth filters are used as they tend to be closely spaced in wavelength and errant zeropoints can create sharp features in the SEDs.  A common procedure to ensure the best-quality $z_{photo}$ is to refine the photometric zeropoints of a catalog using a subsample of galaxies with spectroscopic redshifts \citep[e.g.,][]{Ilbert2006,Ilbert2009,Brammer2011,Whitaker2011}.   
\newline\indent
We perform this process using an iterative zeropoint offset code developed for the NMBS \citep[see][]{Whitaker2011}.  We use only the highest-quality spectroscopic redshifts from zCOSMOS (3 $< CC <$ 5) as well as a set of 19 $z_{spec}$ of massive galaxies at $z >$ 1 from other programs in the COSMOS field \citep{Onodera2012,vandesande2011,vandesande2012,Bezanson2013}.  
\newline\indent
The procedure is as follows.  Firstly, EAZY is run on SEDs of the spectroscopic redshift sample fixing the photometric redshift to the spectroscopic redshift.  The median offset in each filter compared to the best-fit template is measured for the full sample.  The photometry for all galaxies is adjusted by this amount and EAZY is re-run.  The process is repeated until the median offsets converge to a value of less than 0.01 mag in every filter.  When calculating the offsets, the K$_{s}$-band is used as the ``anchor" filter and is not adjusted.  Given that the offsetting is based on colors, having a filter that is not adjusted is important so that the zeropoints do not drift in an absolute sense while the relative zeropoints are being improved.  We note though that the K$_{s}$ zeropoint was initially adjusted by 0.08 mag to bring it into better agreement with other surveys (see Appendix).  We have also not computed offsets for the IRAC and $GALEX$ bands.  It is unclear how well the EAZY templates should reproduce these wavelength ranges and so we chose to not iterate them as it may introduce incorrect colors because of the adopted template set.  
\newline\indent
\begin{deluxetable*}{lll}
\tabletypesize{\footnotesize}
\scriptsize
\tablecaption{Summary of Photometric Redshift Catalog}
\tablehead{\colhead{Column} & \colhead{ Parameter Name } & \colhead{ Description } 
}
\startdata
1 & id & Object identifier number \nl
2 & z\_spec & Spectroscopic Redshift from zCOSMOS (no redshift = -1) \nl
3 & chi & $\chi^2$ of the best-fitting template \nl
4 & z\_peak & Photometric redshift from the peak of the P($z$) distribution \nl
5, 6 & l68, u68 & Upper and lower 68\% confidence intervals on z\_peak \nl
7, 8 & l95, u95 & Upper and lower 95\% confidence intervals on z\_peak \nl
9, 10 & l99, u99 & Upper and lower 99\% confidence intervals on z\_peak \nl
11 & peak\_prob & Peak probability \nl
12 & nfit & Number of filters used to determine z\_peak \nl 
\enddata
\end{deluxetable*}
\begin{figure*}
\plotone{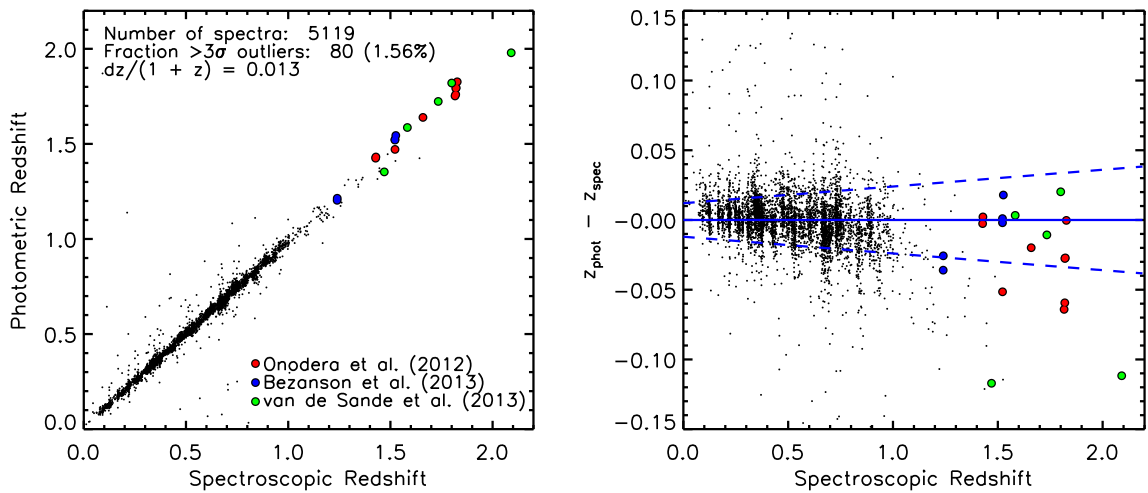}
\caption{\footnotesize Left panel: Photometric redshift from the UltraVISTA K$_{s}$-selected catalog vs. spectroscopic redshift from the zCOSMOS catalog (black dots).  Only galaxies with high-quality spectroscopic redshifts and uncontaminated photometry are shown.  Red galaxies with high-quality spectroscopic redshifts from Onodera et al.~(2012) are plotted as red symbols.  Also plotted are red galaxies with redshifts from \cite{Bezanson2013} and \cite{vandesande2012}.  Right panel: Difference between spectroscopic redshift and photometric redshift as a function of spectroscopic redshift.  The dashed lines show the rms dispersion of $\delta$$z$ = 0.13/(1 + $z$), which is calculated from the zCOSMOS sample once catastrophic outliers are removed.  The Onodera, Bezanson, and van de Sande spectroscopic redshifts suggest that the rms dispersion at $z > $ 1.4 is likely to be larger than that measured from the zCOSMOS sample.  }
\end{figure*}
In Table 3 we list the zeropoint offsets determined for all bands.  These are typically small for the optical broad band filters ($\sim$ 0.05 mag), but can be larger for the optical medium bands (up to 0.17 mag).  Similar size offsets for those bands were also found in the NMBS \citep[see][Table 4]{Whitaker2011}.  Offsets of the order 0.1 -- 0.2 mag are seen for the UltraVISTA bands.  These are comparable to the offsets seen between the UltraVISTA data and the CFHT WIRCAM data by \cite{McCracken2012}.
\subsection{Photometric Redshift Catalog and Comparison with Spectroscopic Redshifts}
In Table 4 we show the layout of the photometric redshift catalog from EAZY.  The parameter z\_peak corresponds to the peak probability of the P($z$) function, and is considered to be the most likely z$_{phot}$.  The 68\% and 95\% confidence intervals are calculated by integrating the P($z$) function.  
\newline\indent
In the left panel of Figure 5 we show a comparison between the $z_{phot}$ and $z_{spec}$ for the zCOSMOS redshifts, as well as the high-redshift spectroscopic samples from \cite{Onodera2012},\cite{Bezanson2013}, and \cite{vandesande2012}.  In general, the photometric redshifts compare well to the spectroscopic redshifts, although we note that this comparison is primarily from galaxies at $z_{spec}$ $<$ 1.5.  Galaxies that are $>$ 3$\sigma$ outliers from the one-to-one relation based on their redshift errors from EAZY are considered as catastrophic outliers.  The fraction of these galaxies is low (1.56\%), and the rms dispersion around the one-to-one relation for the remainder of the sample is 0.013 in $\delta$$z$/(1+$z$).  
\newline\indent
The $z_{phot}$ accuracy compares well to the $z_{phot}$ accuracy for other catalogs in the COSMOS field.  \cite{Ilbert2009} measure a catastrophic outlier fraction of 0.7\%, and $\delta$$z$/(1+$z$) = 0.007.  This is slightly better than for our K$_{s}$-selected catalog; however, \cite{Ilbert2009} use a more restricted set of $z_{spec}$ for this test ($\sim$ 4000 $z_{spec}$ compared to $\sim$ 5000 $z_{spec}$ in our catalog).  \cite{Whitaker2011} find a catastrophic outlier fraction of 5\% with a $\delta$$z$/(1+$z$) = 0.008.  This is a higher outlier fraction but a better rms dispersion.  This difference is mostly likely from our inclusion of a 5\% systematic error in the error bars when fitting with EAZY which reduces catastrophic outliers at the expense of some redshift precision.
\begin{figure*}
\plottwo{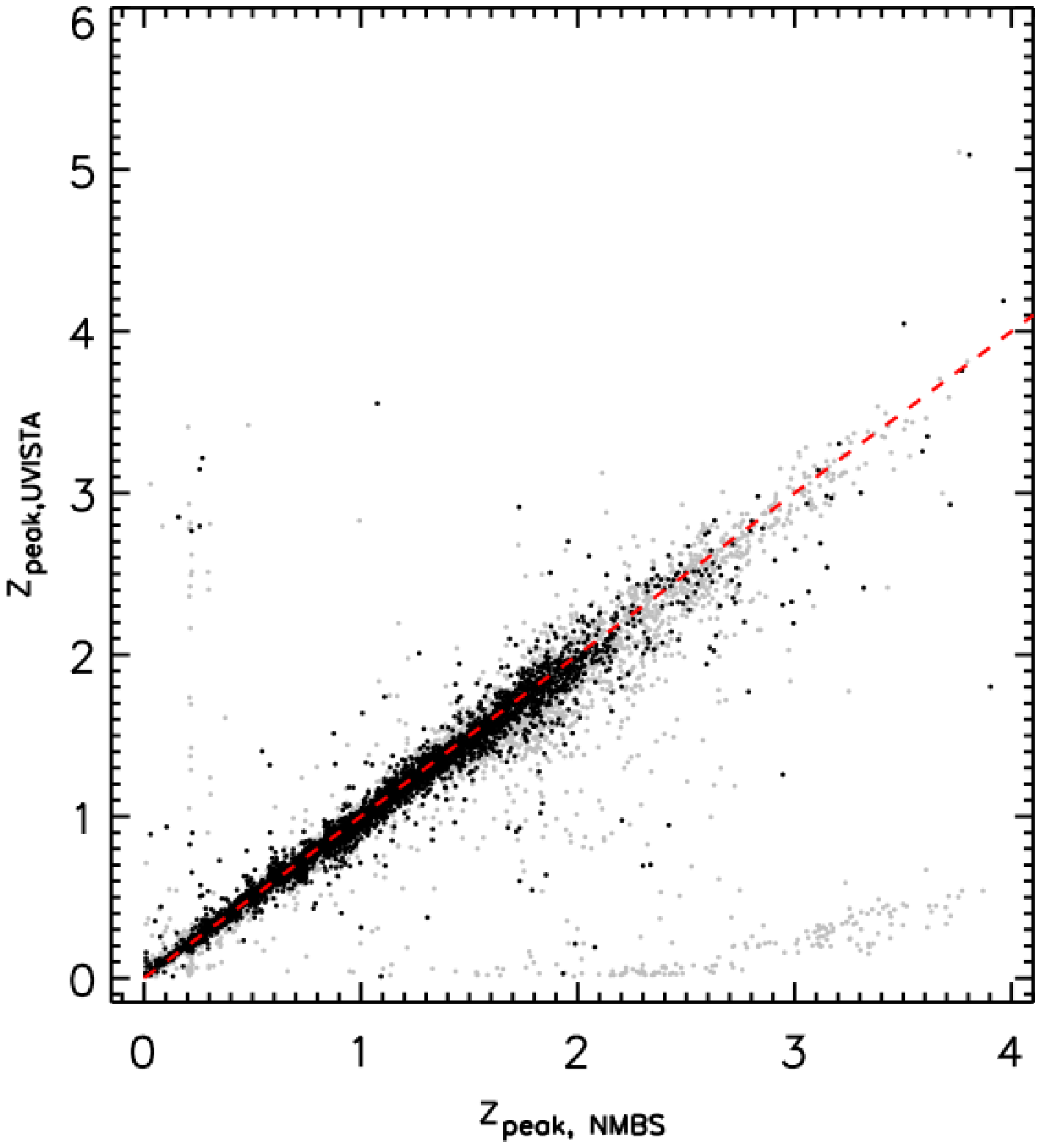}{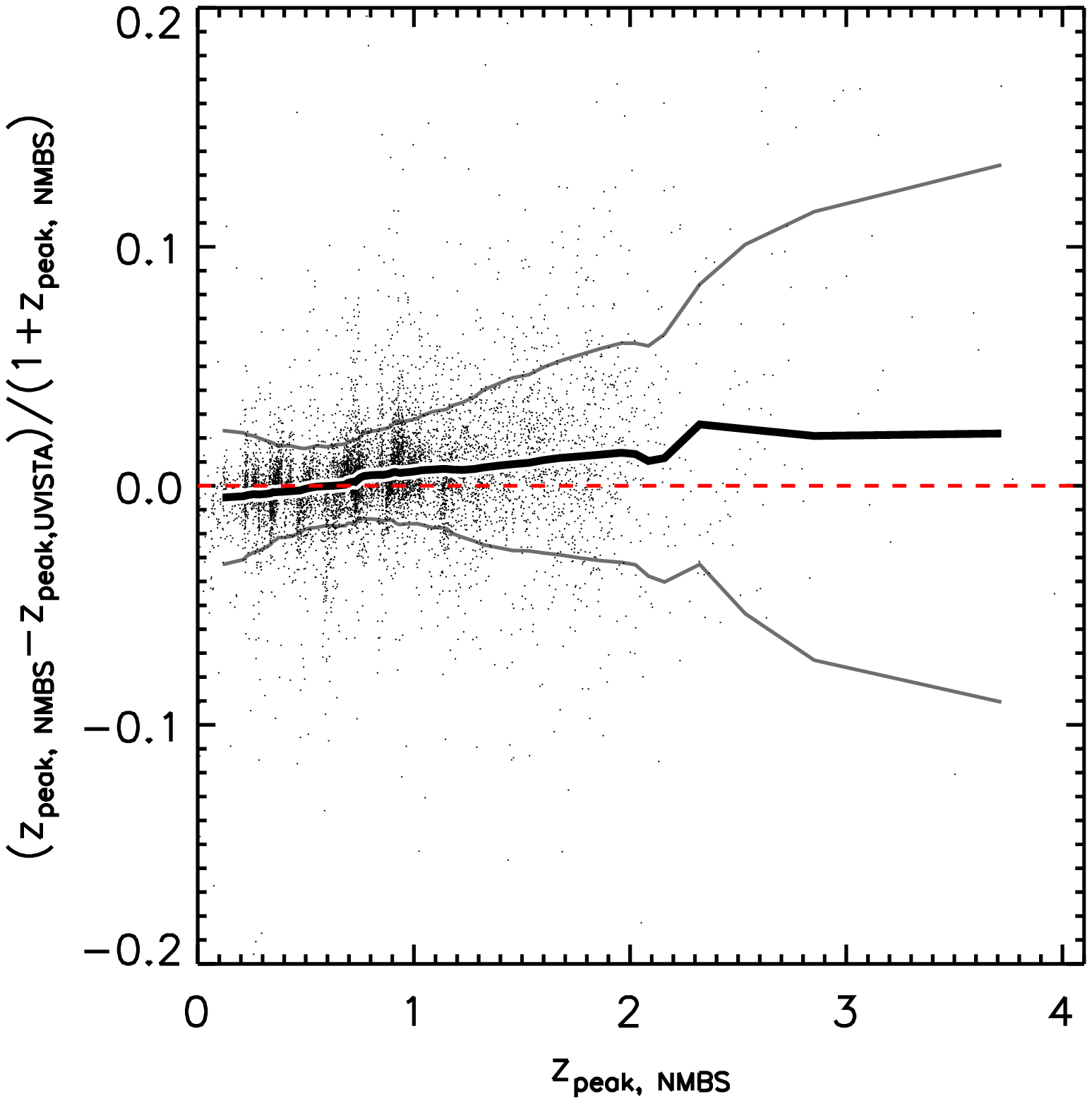}
\caption{\footnotesize Left panel: Photometric redshifts from UltraVISTA vs. photometric redshifts from the NMBS in the overlap region between the two surveys.  Galaxies with stellar masses $>$ 95\% mass-completeness limit for the survey are shown as dark circles, and those below the limit are shown as light circles.  Right panel: Difference in photometric redshift between UltraVISTA and NMBS as a function of photometric redshift.  The solid curve shows the median difference and the gray curves show the rms.  The agreement between the two surveys is extremely good, particularly for the mass-complete sample where the $>$5 $\sigma$ outlier fraction is 2.0\% and the rms is $\delta$$z$/(1+$z$) = 0.026.}
\end{figure*}
\begin{figure}
\plotone{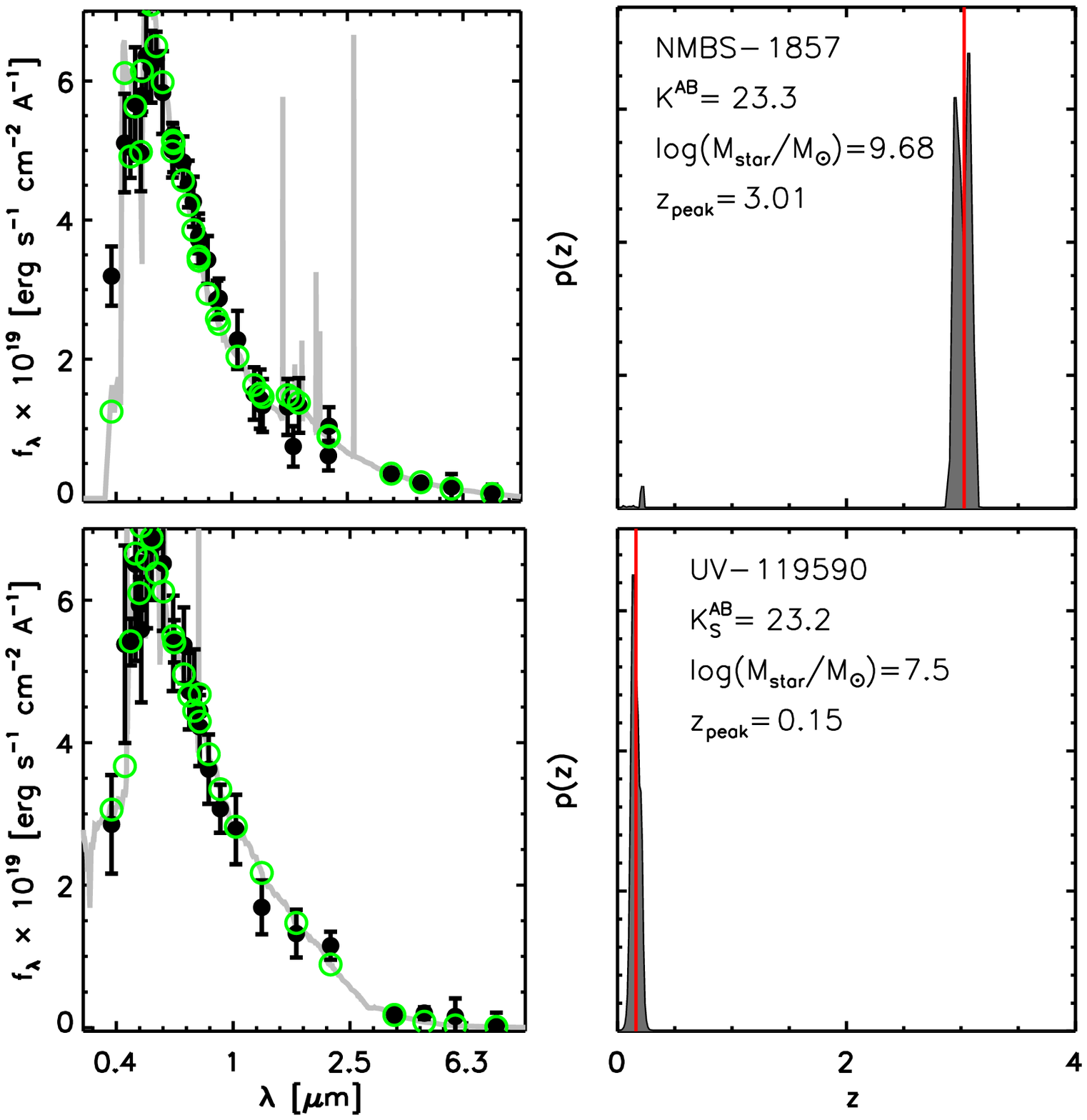}
\caption{\footnotesize Example of a an extreme blue galaxy with a photometric redshift that is very different between the UltraVISTA and NMBS.  The NIR medium bands in NMBS suggest a Balmer break, which is less well-resolved with the UltraVISTA broad bands.  We have corrected the photometric redshift of the extreme blue galaxies at $z <$ 0.5 to their best fit solution at $z >$ 1.5.  The correction is primarily aesthetic, as these galaxies make up $<$ 1\% of all galaxies and are typically well below the mass-completeness limit of the survey (at either redshift).}
\end{figure}
\subsection{Comparison with NMBS Photometric Redshifts}
The zCOSMOS $z_{spec}$ provide a useful diagnostic of the accuracy of the $z_{phot}$; however, there are few $z_{spec}$ at $z >$ 1.5, leaving the accuracy in this redshift range less certain.  The NMBS covers 0.22 deg$^2$ of the COSMOS field and here we compare the $z_{phot}$ determined between each survey.  The NMBS catalog has similar data to the UltraVISTA catalog.  It is also a K$_{s}$-selected catalog and contains photometry from the Subaru broadband and medium band data.  It also uses the same source-fitting code for the $GALEX$ and IRAC photometry and the $z_{phot}$ for the NMBS have been computed using EAZY.  The primary differences are that the NMBS has 6 NIR filters compared to 4 from UltraVISTA.  This NMBS is also $\sim$ 0.6 mag shallower than UltraVISTA.  The NMBS does have deeper optical photometry in the u$^{*}$g$^{\prime}$r$^{\prime}$i$^{\prime}$z$^{\prime}$ bands from the CFHTLS data which are not included in the UltraVISTA catalog.
\newline\indent
In the left panel of Figure 6 we plot the $z_{phot,UVISTA}$ vs. $z_{phot,NMBS}$.  Dark circles are for galaxies that are above the 95\% mass-completeness limit of the survey at a given redshift \citep[see][]{Muzzin2013c}, and light symbols are for those below the 95\% mass-completeness limit.  For the mass-complete sample, the agreement between UltraVISTA and NMBS is excellent.  The fraction of galaxies at 0 $< z <$ 4 that are $>$ 5$\sigma$ outliers is 2.0\%, with a scatter in $\delta$$z$/(1+$z$) = 0.026.  Including the full sample of galaxies the number of $>$ 5$\sigma$ outliers increases to 3.7\%.
\newline\indent
In the right panel of Figure 6 we plot the difference between the $z_{phot}$ as a function of the $z_{photo,NMBS}$.  At $z >$ 1, there is a slight systematic bias between the two surveys of $\delta$$z$/(1+$z$) = 0.005, with the $z_{phot}$ from NMBS being slightly higher for a given galaxy.  This was also suggested by the small number of galaxies in this redshift range with spectroscopic redshifts ($\S$ 4.2).  Small systematics like this are not completely unexpected.  The NMBS used NIR medium bands in order to better-trace the Balmer/4000\AA~break and provide improved $z_{photo}$ for galaxies at $z >$ 1.5 \cite[see][]{Whitaker2011}.  Even with the slightly deeper data available from UltraVISTA, it would be unrealistic to expect NIR broad bands to perform as well as NIR medium bands when determining $z_{phot}$.
\subsection{Correction for Extreme Blue Galaxies} 
\indent
In the left panel of Figure 6, it is clear that a subsample of the fainter galaxies have $z_{phot}$ that do not agree well between the surveys.  These are seen as the line of galaxies that have $z_{phot}$ $<$ 0.5 in UltraVISTA but a range of much higher $z_{phot}$ in NMBS.  There is a complementary population running along the Y-axis; however, it is substantially smaller suggesting that on average the $z_{phot}$ of this population may have been underestimated in the UltraVISTA catalog.  
\newline\indent
As an example, we show the SEDs and P($z$) of one of these objects in both surveys in Figure 7.  Examination of the P($z$) for these galaxies in both surveys shows that they are frequently bi-modal with a peak both at $z <$ 0.5, as well as a peak at $z >$ 1.5.  The SEDs of these galaxies are also typically very blue.  They have a single feature, which is a break starting in the bluest optical bands.  This break is interpreted as the Balmer-break in the low-redshift solution, and as the Lyman break for the high-redshift solution. 
\newline\indent
These galaxies are problematic, but very few of them have spectroscopic redshifts so determining which solution is the correct one is non-trivial.  Several lines of evidence point to the NMBS solution of high-redshift Lyman breaks being the correct ones.  Firstly, many of the galaxies with the high-redshift solution do show weak Balmer breaks in the NIR medium bands (see Figure 7).  These breaks are well traced by the NIR medium bands.  They are also seen in the UltraVISTA broad bands, but are not as well traced, given that there are less filters.  As in Figure 7, the break in UltraVISTA is usually manifest as a single deviant point (either the Y or J band) to an otherwise good fit.  Secondly, the NMBS uses the deep CFHTLS u$^{*}$ data which covers only a portion of the UltraVISTA field.  It also uses as smaller color aperture (1.5$^{\prime\prime}$ compared to 2.1$^{\prime\prime}$), so the photometric errors in the u$^{*}$ band in the NMBS are smaller by a factor of $\sim$ 1.5.  This makes the u$^{*}$ band breaks clearer, and in many cases rules out the possibility of the break being the Balmer break, which is a weaker feature.  Lastly, once rest-frame colors are computed (see $\S$ 5.2), it is clear that many of these galaxies lie in a portion of the UVJ diagram not populated by other galaxies.  
\newline\indent
Given these lines of evidence, it is reasonable to expect that most of these galaxies are likely to be high-redshift galaxies and so we apply a correction to their photometric redshift.  The NMBS only covers a fraction of the UltraVISTA field, so we cannot use those data to improve things.  Instead, we identify the population based on their location in the UVJ diagram.  Galaxies that are much bluer than the overall $z <$ 0.5 population in both U - V and V - J (rest-frame) are selected as the candidate high-$z$ population.  For these galaxies we re-run EAZY, but only allowing solutions in the range 1.5 $< z <$ 6.0.  This prevents a solution at $z <$ 0.5 and forces the $z_{phot}$ to the $z >$ 1.5 solution.  
\newline\indent
While this correction is somewhat {\it ad hoc} we note that it affects only a small fraction of galaxies in the catalog.  Of the total sample of 262 615 sources, only 2415 ($<$ 1\%) are affected by this correction.  More importantly, of this 1\%, the majority are quite faint, and as shown in Figure 6, almost none lie above the mass-completeness limit of the survey no matter whether their solution is at high-redshift or low-redshift.  Because of this, the correction of their $z_{phot}$ does not affect any results based on the mass-complete sample.
\section{SED Modeling and Stellar Population Parameters}
\begin{figure}
\plotone{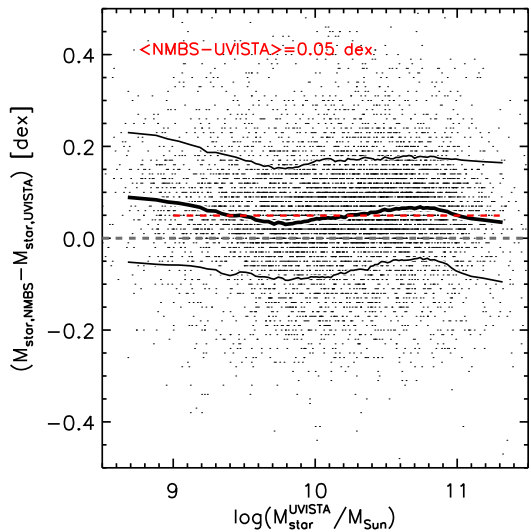}
\caption{\footnotesize Comparison of stellar masses determined in the NMBS and UltraVISTA as a function of stellar mass.  The dark solid curve is the running median and the lighter solid curves encompass the 68-percentile of the distribution  There is a trend for systematically higher masses in the NMBS, but no significant trend with stellar mass.  }
\end{figure}
\begin{figure*}
\plotone{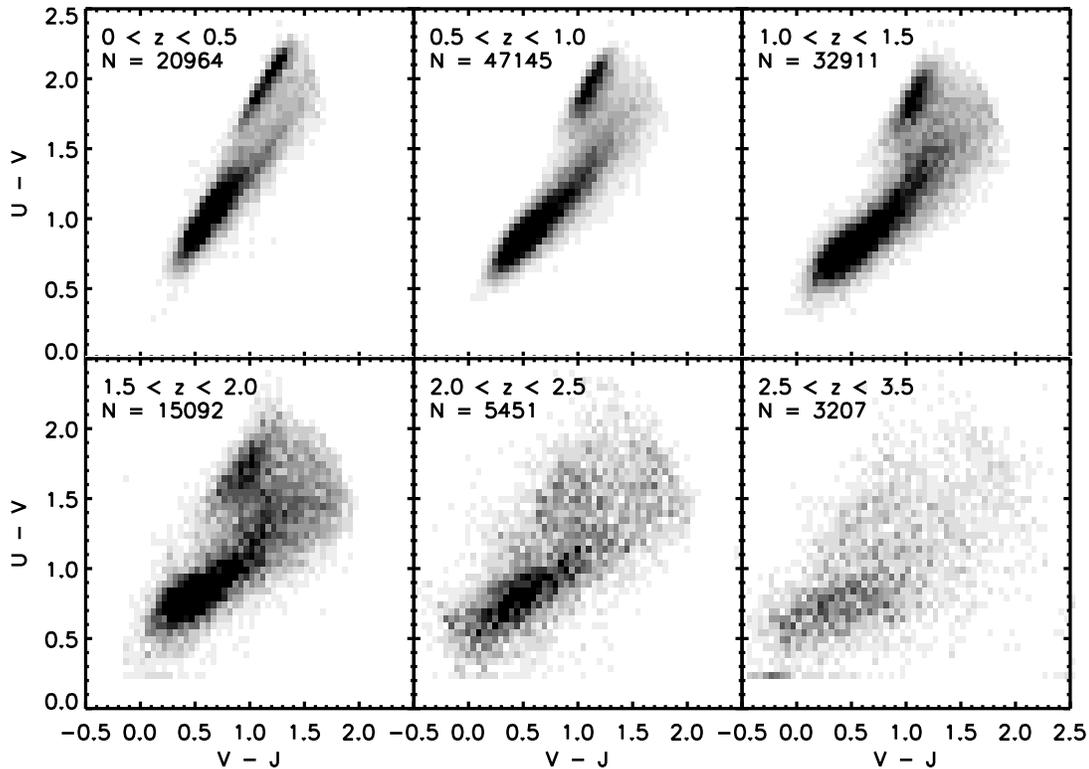}
\caption{\footnotesize The U - V vs. V - J color-color diagram for galaxies with S/N(K$_{s}$) $>$ 7.  The well-known bi-modality between quiescent galaxies and star forming-galaxies can be clearly seen in the galaxy population up to $z \sim$ 2.  Thereafter the bi-modality becomes less pronounced.  Also visible in the figure is the continued reddening of the passive population with decreasing redshift.  This reddening is expected from a passively-evolving population.}
\end{figure*}
\begin{deluxetable*}{lll}
\tabletypesize{\footnotesize}
\scriptsize
\tablecaption{Summary of Stellar Population Parameter Catalog}
\tablewidth{7.0in}
\tablehead{\colhead{Column} & \colhead{ Parameter Name } & \colhead{ Description } 
}
\startdata
1 & id & Object identifier number \nl
2, 3, 4 & z, l68\_z, u68\_z & Photometric redshift and 68\% confidence intervals from EA$z$y \nl
5, 6, 7 & ltau, l68\_ltau, u68\_ltau & Best-fit value of log($\tau$) and 68\% confidence intervals \nl
8, 9, 10 & lage, l68\_lage, u68\_lage & Best-fit value of log($t$) and 68\% confidence intervals \nl
11, 12, 13 & A$_{v}$, l68\_A$_{v}$, u68\_A$_{v}$ & Best-fit value of A$_{v}$ and 68\% confidence intervals \nl
14, 15, 16 & lmass, l68\_lmass, u68\_lmass & Best-fit value of log(M$_{star}$/M${\odot}$) and 68\% confidence intervals \nl
17, 18, 19 & lsfr, l68\_lsfr, u68\_lsfr & Best-fit value of log(sfr) from the SED and 68\% confidence intervals \nl
20, 21, 22 & lssfr, l68\_lssfr, u68\_lssfr & Best-fit value of log(ssfr) from the SED and 68\% confidence intervals \nl
23 & chi & $\chi^2$ of best-fitting model \nl
\enddata
\end{deluxetable*}
\begin{figure*}
\plotone{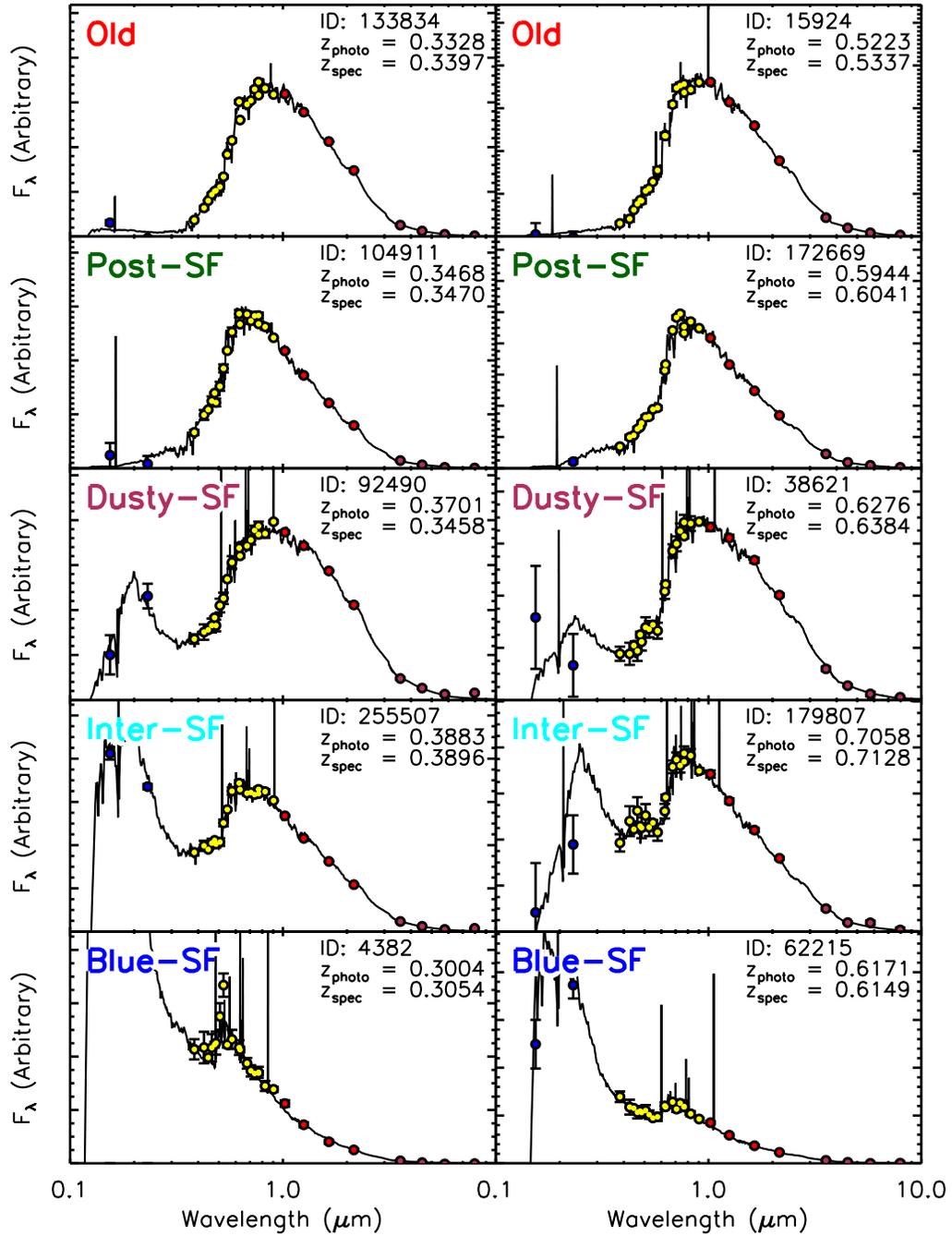}
\caption{\footnotesize Example SEDs for galaxies of different types at 0 $< z < $ 1.0 with S/N(K$_{s}$) $>$ 10 chosen from their location in the UVJ diagram (see text).  In general the fits are very good, and clear features such as the Lyman break, Balmer break, and 4000\AA~break can be seen in the galaxy populations. }
\end{figure*}
\begin{figure*}
\plotone{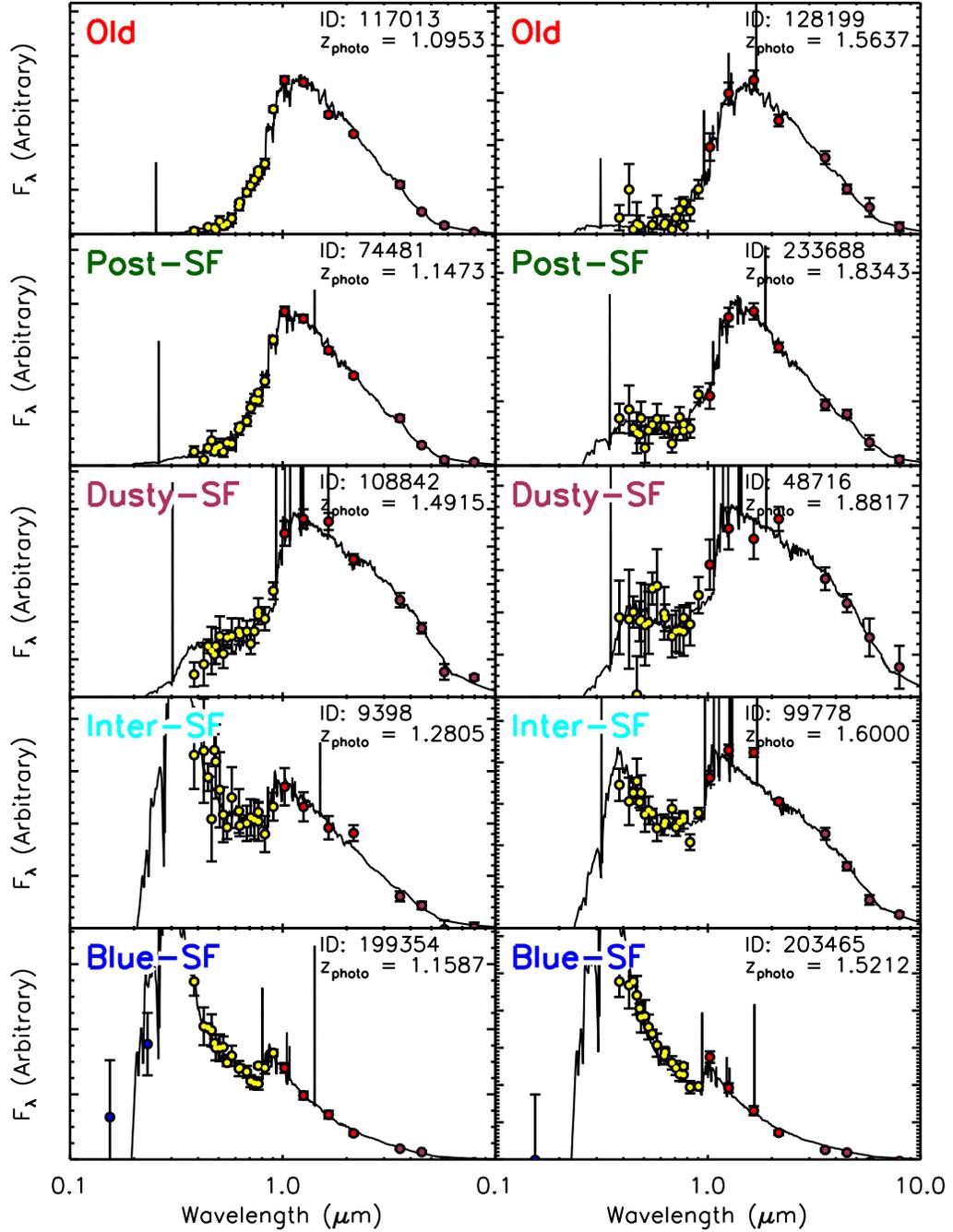}
\caption{\footnotesize As Figure 10, but for galaxies at 1 $< z <$ 2. }
\end{figure*}
\begin{figure*}
\plotone{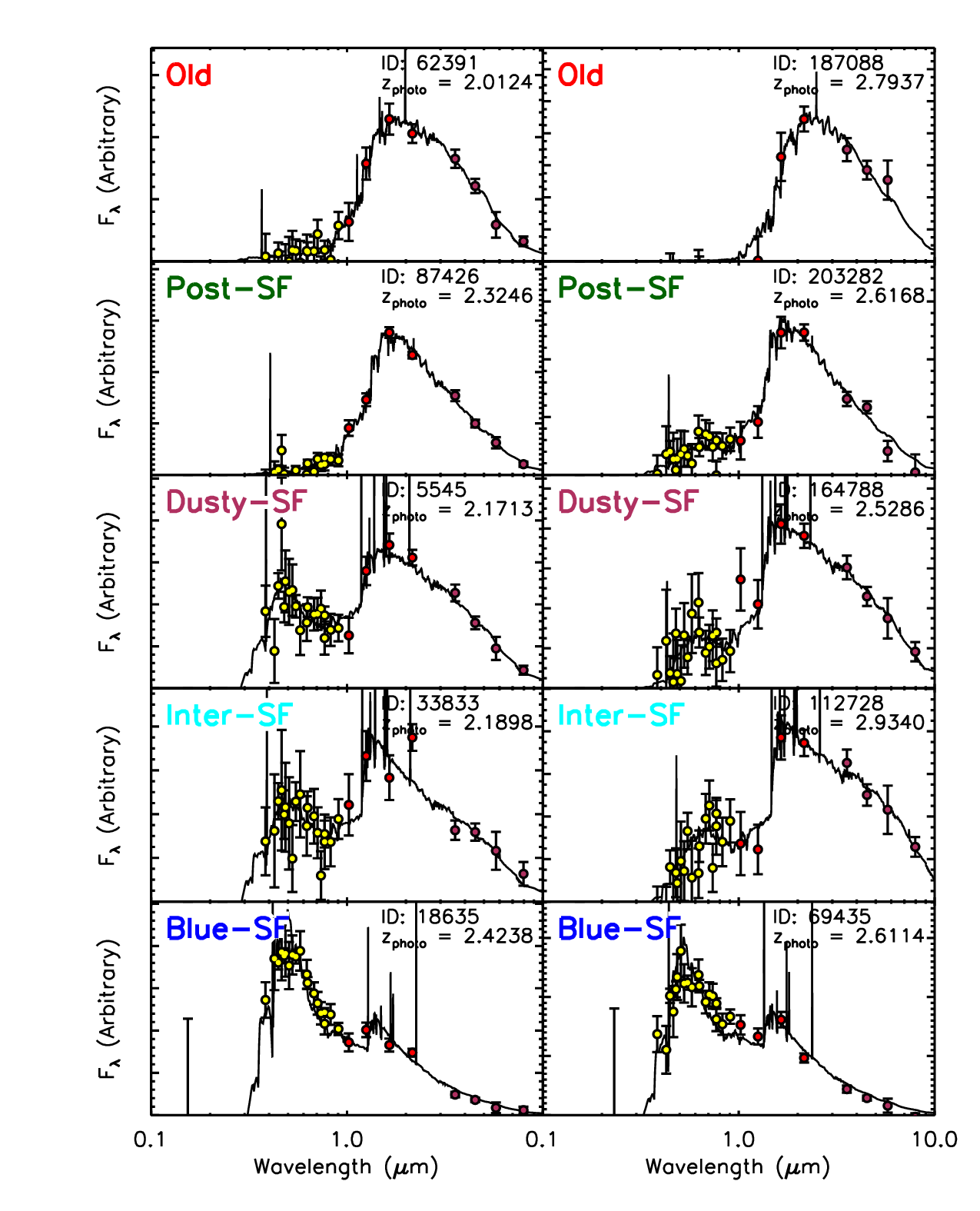}
\caption{\footnotesize As Figure 10, but for galaxies at 2 $< z <$ 3.  }
\end{figure*}
\subsection{Stellar Masses, Ages, and Dust Extinction}
\indent
Stellar population parameters are determined by fitting galaxy SEDs using the FAST code \citep{Kriek2009}.  We provide two catalogs of population parameters for the catalogs, one fit to SEDs generated using the \cite{Bruzual2003} models and one fit to SEDs generated using the \cite{Maraston2005} models.  For both sets of models we assume solar metallicity, a \cite{Chabrier2003} IMF, and a \cite{Calzetti2000} dust extinction law.
\newline\indent
To construct the set of template SEDs, we assume galaxies have exponentially-declining star formation histories (SFHs) of the form SFR $\propto$ exp(-$t$/$\tau$), where $t$ is the time since the onset of star formation and $\tau$ is the $e$-folding star formation timescale in units of Gyr.  We allow Log($\tau$) to vary between 7.0 and 10.0 in increments of 0.2, and Log($t$) to vary between 7.0 and 10.1, in increments of 0.1.  For all galaxies we restrict $t$ to be less than the age of the universe at the redshift of the galaxy.  We also fit for visual attenuation of the galaxies (A$_{v}$) assuming a uniform dust screen geometry and allow A$_{v}$ to vary between 0 and 4.  All galaxies are fit assuming their redshift is the best-fit EAZY $z_{phot}$.  In all, we fit four parameters per galaxy: $\tau$, $t$, A$_{v}$, and a normalization.  The stellar mass (M$_{star}$) is then determined from mass-to-light ratio of the best-fit SED multiplied by the best-fit normalization of the SED.  The layout of the catalogs of best-fit stellar population parameters is summarised in Table 5.
\newline\indent
In Figure 8 we show a comparison between the M$_{star}$ measured for the galaxies in the NMBS catalog vs. the M$_{star}$ measured for galaxies in the UltraVISTA catalog as a function of mass.  In general the agreement is good.  There is a systematic difference of 0.05 dex between the two with galaxies being more massive in the NMBS catalog.  
\subsection{Rest-frame Colors}
The rest-frame U - V vs. V - J diagram has become a popular way to differentiate between star-forming and quiescent galaxies \citep{Labbe2005,Wuyts2007, Williams2009,Brammer2011,Whitaker2011,Patel2012a,Patel2012b}.  This approach is similar to the observed BzK diagram that has been used in the past \citep[e.g.,][and numerous others]{Daddi2005,Daddi2007}, but allows for a cleaner separation of star-forming and quiescent galaxies because colors are defined in the rest-frame rather than observed frame, removing any redshift-dependence of the colors.  
\newline\indent
We calculate rest-frame U - V and V - J colors for all galaxies using EAZY.  EAZY determines the colors by integrating the best-fit SED through the redshifted filter curves over the appropriate wavelength range.  For the U and V filter we use the response curves defined in \cite{MaizApellaniz2006}, and for the J filter we used the 2MASS filter curve from \cite{Skrutskie2006}.  The rest-frame U - V and V - J colors are listed in Table 6.
\begin{deluxetable*}{lll}
\tabletypesize{\footnotesize}
\scriptsize
\tablecaption{Summary of Rest-frame color and UV and IR Luminosity Catalog}
\tablewidth{7.0in}
\tablehead{\colhead{Column} & \colhead{ Parameter Name } & \colhead{ Description } 
}
\startdata
1 & id & Object identifier number \nl
2 & U - V & Rest-frame U - V color \nl
3 & V - J & Rest-frame V - J color \nl
4, 5, 6 & L$_{2800}$ & Total UV luminosity and 68\% confidence intervals \nl
7, 8, 9 & L$_{IR}$ & Total IR luminosity and 68\% confidence intervals \nl
10, 11, 12 & SFR$_{UV,uncorr}$ & UV star formation rate and 68\% confidence intervals, uncorrected for dust extinction \nl
13, 14, 14 & SFR$_{IR}$ & IR star formation rate and 68\% confidence intervals \nl
\enddata
\end{deluxetable*}
\subsection{The UVJ Diagram}
In Figure 9 we plot grayscale histograms of the U - V vs. V - J colors for galaxies with S/N(K$_{s}$) $>$ 7 in various redshift bins between 0 $< z <$ 3.5.  The galaxy population is clearly separated into two clumps in color-color space up to $z =$ 2.  The reddest of the two clumps has colors similar to that of quiescent and passively-evolving galaxies, whereas the bluer clump has a much wider range of colors, usually interpreted as the star-forming population with a range of dust extinctions and geometries \citep[e.g.,][]{Labbe2005,Williams2009,Brammer2009,Patel2012b}.  In general, the UVJ diagram in COSMOS shows the same structure as has been seen in other surveys \citep[e.g.,][]{Williams2009,Whitaker2011}; however, the superior volume in UltraVISTA provides an increase in the number of galaxies by a factor of $\sim$ 4 - 5. 
\newline\indent
An interesting feature of the diagram revealed by the large sample of galaxies in UltraVISTA is the continued reddening of the quiescent population with decreasing redshift.  Whereas the quiescent population is primarily located in a single clump with colors of U - V = 1.7 and V - J = 1.0 at $z =$ 2, it appears as more of a sequence at $z =$ 0.  This reddening with decreasing redshift is precisely what is expected for a passively-evolving population of galaxies \citep[e.g.,][]{Wuyts2007}.  
\newline\indent
It is tempting to make further inferences on the evolution of the galaxy population using Figure 9.  Doing so requires a more quantitative analysis, in particular it is important to adopt the appropriate limits in M$_{star}$ with redshift in order to define a mass-complete sample.  Figure 9 is meant to be illustrative and is made with a S/N cut, not a M$_{star}$-cut, so it becomes increasingly skewed to lower-mass galaxies at lower-redshift.  A full analysis of the UVJ diagram using the appropriate mass limits will be presented in a future paper.
\subsection{Example SEDs}
As an example of the quality of the SEDs and SED fitting, we show examples of the observed galaxy SEDs and the best-fit EAZY SEDs from the catalog in Figures 10 - 12.  At each redshift, SEDs classified by their location in the UVJ diagram are shown in order to demonstrate the SEDs for a range of different galaxy types.  We chose a total of five different SED classifications, two types of SEDs in the quiescent population, and three types in the star-forming population.  Within the quiescent population we chose those that are located near the red-tip in U - V vs. V - J (labelled ``old"), and those that are near the blue-tip in  U - V vs. V - J, but still have quiescent colors.  These blue-tip galaxies have colors that are similar to recently quenched galaxies, or ``post star formation galaxies" \citep[e.g.,][]{Kriek2010,Whitaker2012b} and are labelled ``post-SF".  
\newline\indent
For the star forming population, we divided the sequence of galaxies going from blue U - V and blue V - J colors, to red U - V and red V - J colors into three bins.  The three regions cover the bluest population, an intermediate population, and the reddest (dustiest) population (labeled blue-SF, inter-SF, and dusty-SF, respectively).  
\newline\indent
The galaxies in each bin are chosen to have S/N(K$_{s}$) $>$ 10, but are otherwise selected at random so that they are representative of typical SEDs for that SED type and redshift range.  The only exception to this is that for the lowest redshift bin we use only galaxies that have $z_{spec}$ from zCOSMOS in order to show the agreement between photometric and spectroscopic SEDs.
\newline\indent
Figures 10 - 12 demonstrate the excellent quality of the SEDs and SED fitting, all the way up to $z$ = 2.5.  They also show how at $z <$ 1 the optical medium bands trace the Balmer and 4000\AA~break extremely well, and why such good quality $z_{phot}$ can be determined.  At $z > 1.5$ the UltraVISTA bands trace the break.  In particular, it can be seen how the Y-band data is extremely useful for connecting what would be a large gap in wavelength space between the z$^{\prime}$-band and J-band.  At $z >$ 2.5 the IRAC data becomes increasingly important, because these are the only filters that remain redward of the break.  The best-fit SEDs for all galaxies using EAZY and and FAST for both BC03 and M05 models are available for download with the catalog.
\begin{figure*}
\plotone{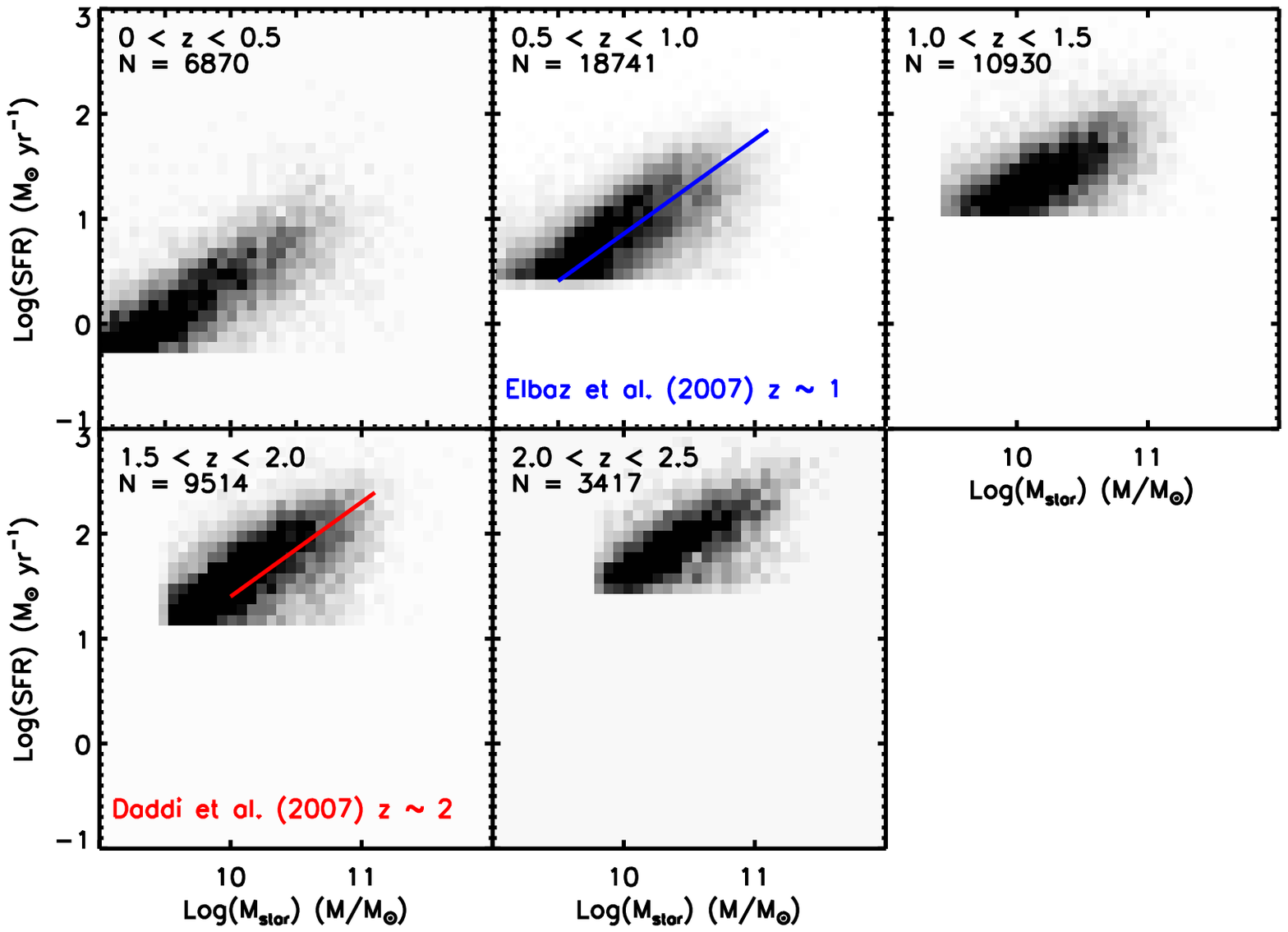}
\caption{\footnotesize Log(SFR) vs. M$_{star}$ in difference redshift bins for star forming galaxies selected using the UVJ diagram (Figure 9).  As in many previous works there is a correlation between the SFR of galaxies and their mass.  The blue and red curves are power-law fits to the relations determined by \cite{Elbaz2007} and \cite{Daddi2007} using data from the GOODS survey.  The normalization and slope of the star formation main sequence evolves similarly between the two studies.  }
\end{figure*}
\subsection{UV and Infrared Luminosities and SFRs}
The SED fitting to the BC03 and M05 models results in an estimated SFR for each galaxy (see Table 5).  Those values can be considered total, dust-corrected SFRs (primarily constrained by the UV flux).  They may be indicative; however, we caution against using those values in a quantitative way.  This is because the estimated SFR is strongly influenced by the assumption of an exponentially-declining SFH.  If this is not the correct SFH \citep[which is likely to be true for many star forming galaxies, e.g.,][]{Maraston2010,Papovich2011}, then they can be substantially in error.  
\newline\indent
A better measure of the instantaneous SFR comes from the rest-frame UV flux (L$_{2800}$) and the rest-frame total infrared luminosity (L$_{IR}$).  With the availability of rest-frame UV information (from $GALEX$ to optical) as well as rest-frame MIR data (from MIPS-24$\micron$) we can calculate these quantities, or upper limits on them, for all galaxies.  These can then be converted to SFRs based on the standard conversion factors \citep[e.g.,][]{Kennicutt1998}. 
\newline\indent
To determine L$_{2800}$ we use EAZY.  Similar to the rest-frame colors ($\S$5.2), EAZY integrates the best-fit template over the wavelength range 2600\AA~-- 2950\AA~to determine an L$_{2800}$ for all galaxies.  Because there is $GALEX$ data, this quantity is constrained by data over the full redshift range of the galaxy population.  The L$_{2800}$ values are listed in Table 6.  
\newline\indent
To determine the L$_{IR}$, we extrapolate the measured 24$\micron$ flux using templates.  The \cite{Chary2001,Dale2002} models are based on local galaxy templates calibrated using IRAS and implicitly assume a correlation between L$_{IR}$ and the dust temperature (T$_{d}$).  As has been discussed in recent papers, it appears that this correlation may not hold up to the highest redshifts \citep[e.g.,][]{Muzzin2010,Elbaz2010}.  Instead of using the luminosity-dependent templates, we use a single template to determine L$_{IR}$ for all galaxies.  This approach has been advocated by many recent studies \citep[e.g.,][]{Wuyts2008,Muzzin2010,Elbaz2010}.  We use the log-average of the \cite{Dale2002} templates for this computation \citep[see,][]{Wuyts2008}, but note that using the log-average of the \cite{Chary2001} templates would provide very similar results \citep{Muzzin2010}. We list the L$_{IR}$ determined using this method in Table 6.  
\newline\indent
We convert the L$_{2800}$ into a SFR$_{UV,uncorr}$ using the conversion factor SFR$_{UV,uncorr}$ = 3.234 $\times$ 10$^{-10}$ L$_{2800}$ from \cite{Kennicutt1998}, adapted to a Kroupa IMF by \cite{Bell2005}.  We note that this is the observed SFR, and is not corrected for dust-extinction.  The L$_{IR}$ is converted into a SFR$_{IR}$ using SFR$_{IR}$ = 0.98 $\times$ 10$^{-10}$ L$_{IR}$ from \cite{Kennicutt1998}, adapted to a Kroupa IMF by \cite{Bell2005}.  The total SFR of the galaxy can then be determined via SFR$_{tot}$ = SFR$_{UV,uncorr}$ + SFR$_{IR}$.  These values are also listed in Table 6.
\subsection{The Star Formation Main Sequence}
Many recent studies have shown evidence for a correlation between the SFR of star forming galaxies and their M$_{star}$ \citep[e.g.,][]{Noeske2007,Elbaz2007,Daddi2007,Wuyts2011,Whitaker2012}.  This correlation has become known as the ``star formation main sequence".  In Figure 13 we plot greyscale histograms of the star formation main sequence for star forming galaxies in the UltraVISTA catalog between 0.0 $< z <$ 2.5.  Star forming galaxies have been selected by their location in the UVJ diagram using the prescriptions defined in \cite{Muzzin2013c}, which are based on those determined by \cite{Williams2009}.  The SFRs plotted in Figure 13 are total SFRs which have been computed from SFR$_{tot}$ = SFR$_{UV,uncorr}$ + SFR$_{IR}$.  Only sources that have a $>$ 3$\sigma$ detection at 24$\micron$ are shown.  
\newline\indent
Figure 13 shows clearly the existence of a main sequence of star formation for galaxies in the UltraVISTA catalog.  As a comparison we plot the power-law fits to the main sequence at $z \sim$ 1 and $z \sim$ 2 measured by \cite{Elbaz2007} and \cite{Daddi2007}, respectively using the GOODS data.  There is reasonable agreement between the evolution of the normalization of the relations between the two datasets.  There may be slightly different slopes and normalizations, but this may be because the redshift ranges shown in Figure 13 are not exact matches to those in \cite{Elbaz2007} and \cite{Daddi2007}.  A more detailed look at the star formation main sequence from UltraVISTA will be presented in a future paper.
\section{Conclusion}
In this paper we have presented a public K$_{s}$-selected catalog covering 1.62 deg$^2$ of the COSMOS/UltraVISTA field.  The photometric catalog consists of PSF-matched photometry in 30 bands and covers the wavelength range 0.15$\micron$ $\rightarrow$ 24$\micron$.  The catalog contains 262 615 sources down to the 3$\sigma$ limit of K$_{s}$(2.1$^{\prime\prime}$) $<$ 24.35, and 179 291 sources down to the 90\% completeness limit K$_{s,tot}$ $<$ 23.4.  
\newline\indent
Photometric redshifts have been computed for all galaxies using the EAZY photometric redshift code.  Comparison of the $z_{phot}$ with $z_{spec}$ from zCOSMOS and other spectroscopic surveys shows that at $z <$ 1.5 the catastrophic outlier fraction is low (1.56\%), and the rms scatter is low ($\delta$$z$/(1+$z$) = 0.013.  The $z_{phot}$ also agree well with $z_{phot}$ determined using the NMBS. 
\newline\indent
Included with the catalog is a set of M$_{star}$ and stellar population parameters computed using the FAST SED fitting code.  These M$_{star}$ show good agreement with those calculated in the NMBS, with only a 0.05 dex systematic difference.  Rest-frame U - V and V - J colors are computed for all galaxies.  The population of galaxies in the COSMOS/UltraVISTA field shows a clear bi-modality in color-color space up to $z \sim$ 2, and thereafter the bi-modality begins to break down.  
\newline\indent
The catalog contains measures of L$_{2800}$ and L$_{IR}$ as well the inferred SFRs from those parameters.  Plotting these against the M$_{star}$ for UVJ-selected star forming galaxies shows that there is a star formation main sequence up to $z \sim$ 2.5.  The evolution of the main sequence is consistent with previous measurements. 
\newline\indent
In a companion paper to this one \citep{Muzzin2013c} we show the evolution of the stellar mass functions of galaxies up to $z \sim$ 4 using the K$_{s}$-selected UltraVISTA catalog.  The photometry, photometric redshifts, stellar population parameters, rest-frame colors, UV and IR SFRs from the K$_{s}$-selected catalog are now made available to the astronomical community though the catalog website\footnote{http://www.strw.leidenuniv.nl/galaxyevolution/ULTRAVISTA/}.  We hope that it proves to be a useful resource for further galaxy evolution studies.
\acknowledgements
DM acknowledges support from Tufts University Mellon Research Fellowship in Arts and Sciences.  BMJ and JPUF acknowledge support from the ERC-StG grant EGGS-278202.  The Dark Cosmology Centre is funded by the Danish National Research  Foundation.  JSD acknowledges the support of the European Research Council through the award of an Advanced Grant, and the support of the Royal Society via a Wolfson Research Merit Award.  AM, DM, and MS thank Gigi Guzzo and the Osservatorio Astronomico di Milano-Merate for the hospitality while working on portions of this research.
\bibliographystyle{apj}
\bibliography{apj-jour,myrefs}

\begin{thebibliography}{88}
\expandafter\ifx\csname natexlab\endcsname\relax\def\natexlab#1{#1}\fi

\bibitem[{{Aretxaga} {et~al.}(2011){Aretxaga}, {Wilson}, {Aguilar}, {Alberts},
  {Scott}, {Scoville}, {Yun}, {Austermann}, {Downes}, {Ezawa}, {Hatsukade},
  {Hughes}, {Kawabe}, {Kohno}, {Oshima}, {Perera}, {Tamura}, \&
  {Zeballos}}]{Aretxaga2011}
{Aretxaga}, I., {et~al.} 2011, \mnras, 415, 3831

\bibitem[{{Arnouts} {et~al.}(2007){Arnouts}, {Walcher}, {Le F{\`e}vre},
  {Zamorani}, {Ilbert}, {Le Brun}, {Pozzetti}, {Bardelli}, {Tresse}, {Zucca},
  {Charlot}, {Lamareille}, {McCracken}, {Bolzonella}, {Iovino}, {Lonsdale},
  {Polletta}, {Surace}, {Bottini}, {Garilli}, {Maccagni}, {Picat},
  {Scaramella}, {Scodeggio}, {Vettolani}, {Zanichelli}, {Adami}, {Cappi},
  {Ciliegi}, {Contini}, {de la Torre}, {Foucaud}, {Franzetti}, {Gavignaud},
  {Guzzo}, {Marano}, {Marinoni}, {Mazure}, {Meneux}, {Merighi}, {Paltani},
  {Pell{\`o}}, {Pollo}, {Radovich}, {Temporin}, \& {Vergani}}]{Arnouts2007}
{Arnouts}, S., {et~al.} 2007, \aap, 476, 137

\bibitem[{{Bell} {et~al.}(2005){Bell}, {Papovich}, {Wolf}, {Le Floc'h},
  {Caldwell}, {Barden}, {Egami}, {McIntosh}, {Meisenheimer},
  {P{\'e}rez-Gonz{\'a}lez}, {Rieke}, {Rieke}, {Rigby}, \& {Rix}}]{Bell2005}
{Bell}, E.~F., {et~al.} 2005, \apj, 625, 23

\bibitem[{{Bertin} \& {Arnouts}(1996)}]{Bertin1996}
{Bertin}, E., \& {Arnouts}, S. 1996, \aaps, 117, 393

\bibitem[{{Bezanson} {et~al.}(2013){Bezanson}, {van Dokkum}, {van de Sande},
  {Franx}, \& {Kriek}}]{Bezanson2013}
{Bezanson}, R., {van Dokkum}, P., {van de Sande}, J., {Franx}, M., \& {Kriek},
  M. 2013, \apjl, 764, L8

\bibitem[{{Bielby} {et~al.}(2012){Bielby}, {Hudelot}, {McCracken}, {Ilbert},
  {Daddi}, {Le F{\`e}vre}, {Gonzalez-Perez}, {Kneib}, {Marmo}, {Mellier},
  {Salvato}, {Sanders}, \& {Willott}}]{Bielby2012}
{Bielby}, R., {et~al.} 2012, \aap, 545, A23

\bibitem[{{Bower} {et~al.}(2012){Bower}, {Benson}, \& {Crain}}]{Bower2012}
{Bower}, R.~G., {Benson}, A.~J., \& {Crain}, R.~A. 2012, \mnras, 422, 2816

\bibitem[{{Brammer} {et~al.}(2008){Brammer}, {van Dokkum}, \&
  {Coppi}}]{Brammer2008}
{Brammer}, G.~B., {van Dokkum}, P.~G., \& {Coppi}, P. 2008, \apj, 686, 1503

\bibitem[{{Brammer} {et~al.}(2009){Brammer}, {Whitaker}, {van Dokkum},
  {Marchesini}, {Labb{\'e}}, {Franx}, {Kriek}, {Quadri}, {Illingworth}, {Lee},
  {Muzzin}, \& {Rudnick}}]{Brammer2009}
{Brammer}, G.~B., {et~al.} 2009, \apjl, 706, L173

\bibitem[{{Brammer} {et~al.}(2011){Brammer}, {Whitaker}, {van Dokkum},
  {Marchesini}, {Franx}, {Kriek}, {Labb{\'e}}, {Lee}, {Muzzin}, {Quadri},
  {Rudnick}, \& {Williams}}]{Brammer2011}
---. 2011, \apj, 739, 24

\bibitem[{{Brammer} {et~al.}(2012){Brammer}, {van Dokkum}, {Franx},
  {Fumagalli}, {Patel}, {Rix}, {Skelton}, {Kriek}, {Nelson}, {Schmidt},
  {Bezanson}, {da Cunha}, {Erb}, {Fan}, {F{\"o}rster Schreiber}, {Illingworth},
  {Labb{\'e}}, {Leja}, {Lundgren}, {Magee}, {Marchesini}, {McCarthy},
  {Momcheva}, {Muzzin}, {Quadri}, {Steidel}, {Tal}, {Wake}, {Whitaker}, \&
  {Williams}}]{Brammer2012b}
---. 2012, \apjs, 200, 13

\bibitem[{{Bruzual} \& {Charlot}(2003)}]{Bruzual2003}
{Bruzual}, G., \& {Charlot}, S. 2003, \mnras, 344, 1000

\bibitem[{{Bundy} {et~al.}(2006){Bundy}, {Ellis}, {Conselice}, {Taylor},
  {Cooper}, {Willmer}, {Weiner}, {Coil}, {Noeske}, \& {Eisenhardt}}]{Bundy2006}
{Bundy}, K., {et~al.} 2006, \apj, 651, 120

\bibitem[{{Calzetti} {et~al.}(2000){Calzetti}, {Armus}, {Bohlin}, {Kinney},
  {Koornneef}, \& {Storchi-Bergmann}}]{Calzetti2000}
{Calzetti}, D., {Armus}, L., {Bohlin}, R.~C., {Kinney}, A.~L., {Koornneef}, J.,
  \& {Storchi-Bergmann}, T. 2000, \apj, 533, 682

\bibitem[{{Capak} {et~al.}(2007){Capak}, {Aussel}, {Ajiki}, {McCracken},
  {Mobasher}, {Scoville}, {Shopbell}, {Taniguchi}, {Thompson}, {Tribiano},
  {Sasaki}, {Blain}, {Brusa}, {Carilli}, {Comastri}, {Carollo}, {Cassata},
  {Colbert}, {Ellis}, {Elvis}, {Giavalisco}, {Green}, {Guzzo}, {Hasinger},
  {Ilbert}, {Impey}, {Jahnke}, {Kartaltepe}, {Kneib}, {Koda}, {Koekemoer},
  {Komiyama}, {Leauthaud}, {Le Fevre}, {Lilly}, {Liu}, {Massey}, {Miyazaki},
  {Murayama}, {Nagao}, {Peacock}, {Pickles}, {Porciani}, {Renzini}, {Rhodes},
  {Rich}, {Salvato}, {Sanders}, {Scarlata}, {Schiminovich}, {Schinnerer},
  {Scodeggio}, {Sheth}, {Shioya}, {Tasca}, {Taylor}, {Yan}, \&
  {Zamorani}}]{Capak2007}
{Capak}, P., {et~al.} 2007, \apjs, 172, 99

\bibitem[{{Caputi} {et~al.}(2011){Caputi}, {Cirasuolo}, {Dunlop}, {McLure},
  {Farrah}, \& {Almaini}}]{Caputi2011}
{Caputi}, K.~I., {Cirasuolo}, M., {Dunlop}, J.~S., {McLure}, R.~J., {Farrah},
  D., \& {Almaini}, O. 2011, \mnras, 413, 162

\bibitem[{{Chabrier}(2003)}]{Chabrier2003}
{Chabrier}, G. 2003, \pasp, 115, 763

\bibitem[{{Chary} \& {Elbaz}(2001)}]{Chary2001}
{Chary}, R., \& {Elbaz}, D. 2001, \apj, 556, 562

\bibitem[{{Cirasuolo} {et~al.}(2010){Cirasuolo}, {McLure}, {Dunlop}, {Almaini},
  {Foucaud}, \& {Simpson}}]{Cirasuolo2010}
{Cirasuolo}, M., {McLure}, R.~J., {Dunlop}, J.~S., {Almaini}, O., {Foucaud},
  S., \& {Simpson}, C. 2010, \mnras, 401, 1166

\bibitem[{{Daddi} {et~al.}(2005){Daddi}, {Renzini}, {Pirzkal}, {Cimatti},
  {Malhotra}, {Stiavelli}, {Xu}, {Pasquali}, {Rhoads}, {Brusa}, {di Serego
  Alighieri}, {Ferguson}, {Koekemoer}, {Moustakas}, {Panagia}, \&
  {Windhorst}}]{Daddi2005}
{Daddi}, E., {et~al.} 2005, \apj, 626, 680

\bibitem[{{Daddi} {et~al.}(2007){Daddi}, {Dickinson}, {Morrison}, {Chary},
  {Cimatti}, {Elbaz}, {Frayer}, {Renzini}, {Pope}, {Alexander}, {Bauer},
  {Giavalisco}, {Huynh}, {Kurk}, \& {Mignoli}}]{Daddi2007}
---. 2007, \apj, 670, 156

\bibitem[{{Dale} \& {Helou}(2002)}]{Dale2002}
{Dale}, D.~A., \& {Helou}, G. 2002, \apj, 576, 159

\bibitem[{{Elbaz} {et~al.}(2007){Elbaz}, {Daddi}, {Le Borgne}, {Dickinson},
  {Alexander}, {Chary}, {Starck}, {Brandt}, {Kitzbichler}, {MacDonald},
  {Nonino}, {Popesso}, {Stern}, \& {Vanzella}}]{Elbaz2007}
{Elbaz}, D., {et~al.} 2007, \aap, 468, 33

\bibitem[{{Elbaz} {et~al.}(2010){Elbaz}, {Hwang}, {Magnelli}, {Daddi},
  {Aussel}, {Altieri}, {Amblard}, {Andreani}, {Arumugam}, {Auld}, {Babbedge},
  {Berta}, {Blain}, {Bock}, {Bongiovanni}, {Boselli}, {Buat}, {Burgarella},
  {Castro-Rodriguez}, {Cava}, {Cepa}, {Chanial}, {Chary}, {Cimatti},
  {Clements}, {Conley}, {Conversi}, {Cooray}, {Dickinson}, {Dominguez},
  {Dowell}, {Dunlop}, {Dwek}, {Eales}, {Farrah}, {F{\"o}rster Schreiber},
  {Fox}, {Franceschini}, {Gear}, {Genzel}, {Glenn}, {Griffin}, {Gruppioni},
  {Halpern}, {Hatziminaoglou}, {Ibar}, {Isaak}, {Ivison}, {Lagache}, {Le
  Borgne}, {Le Floc'h}, {Levenson}, {Lu}, {Lutz}, {Madden}, {Maffei}, {Magdis},
  {Mainetti}, {Maiolino}, {Marchetti}, {Mortier}, {Nguyen}, {Nordon},
  {O'Halloran}, {Okumura}, {Oliver}, {Omont}, {Page}, {Panuzzo},
  {Papageorgiou}, {Pearson}, {Perez Fournon}, {P{\'e}rez Garc{\'{\i}}a},
  {Poglitsch}, {Pohlen}, {Popesso}, {Pozzi}, {Rawlings}, {Rigopoulou},
  {Riguccini}, {Rizzo}, {Rodighiero}, {Roseboom}, {Rowan-Robinson},
  {Saintonge}, {Sanchez Portal}, {Santini}, {Sauvage}, {Schulz}, {Scott},
  {Seymour}, {Shao}, {Shupe}, {Smith}, {Stevens}, {Sturm}, {Symeonidis},
  {Tacconi}, {Trichas}, {Tugwell}, {Vaccari}, {Valtchanov}, {Vieira},
  {Vigroux}, {Wang}, {Ward}, {Wright}, {Xu}, \& {Zemcov}}]{Elbaz2010}
---. 2010, \aap, 518, L29

\bibitem[{{Elvis} {et~al.}(2009){Elvis}, {Civano}, {Vignali}, {Puccetti},
  {Fiore}, {Cappelluti}, {Aldcroft}, {Fruscione}, {Zamorani}, {Comastri},
  {Brusa}, {Gilli}, {Miyaji}, {Damiani}, {Koekemoer}, {Finoguenov}, {Brunner},
  {Urry}, {Silverman}, {Mainieri}, {Hasinger}, {Griffiths}, {Carollo}, {Hao},
  {Guzzo}, {Blain}, {Calzetti}, {Carilli}, {Capak}, {Ettori}, {Fabbiano},
  {Impey}, {Lilly}, {Mobasher}, {Rich}, {Salvato}, {Sanders}, {Schinnerer},
  {Scoville}, {Shopbell}, {Taylor}, {Taniguchi}, \& {Volonteri}}]{Elvis2009}
{Elvis}, M., {et~al.} 2009, \apjs, 184, 158

\bibitem[{{Fioc} \& {Rocca-Volmerange}(1999)}]{Fioc1999}
{Fioc}, M., \& {Rocca-Volmerange}, B. 1999, ArXiv Astrophysics e-prints

\bibitem[{{Fontanot} {et~al.}(2009){Fontanot}, {De Lucia}, {Monaco},
  {Somerville}, \& {Santini}}]{Fontanot2009}
{Fontanot}, F., {De Lucia}, G., {Monaco}, P., {Somerville}, R.~S., \&
  {Santini}, P. 2009, \mnras, 397, 1776

\bibitem[{{F{\"o}rster Schreiber} {et~al.}(2004){F{\"o}rster Schreiber}, {van
  Dokkum}, {Franx}, {Labb{\'e}}, {Rudnick}, {Daddi}, {Illingworth}, {Kriek},
  {Moorwood}, {Rix}, {R{\"o}ttgering}, {Trujillo}, {van der Werf}, {van
  Starkenburg}, \& {Wuyts}}]{ForsterSchreiber2004}
{F{\"o}rster Schreiber}, N.~M., {et~al.} 2004, \apj, 616, 40

\bibitem[{{Franx} {et~al.}(2003){Franx}, {Labb{\'e}}, {Rudnick}, {van Dokkum},
  {Daddi}, {F{\"o}rster Schreiber}, {Moorwood}, {Rix}, {R{\"o}ttgering}, {van
  der Wel}, {van der Werf}, \& {van Starkenburg}}]{Franx2003}
{Franx}, M., {et~al.} 2003, \apjl, 587, L79

\bibitem[{{Frayer} {et~al.}(2009){Frayer}, {Sanders}, {Surace}, {Aussel},
  {Salvato}, {Le Floc'h}, {Huynh}, {Scoville}, {Afonso-Luis}, {Bhattacharya},
  {Capak}, {Fadda}, {Fu}, {Helou}, {Ilbert}, {Kartaltepe}, {Koekemoer}, {Lee},
  {Murphy}, {Sargent}, {Schinnerer}, {Sheth}, {Shopbell}, {Shupe}, \&
  {Yan}}]{Frayer2009}
{Frayer}, D.~T., {et~al.} 2009, \aj, 138, 1261

\bibitem[{{Grogin} {et~al.}(2011){Grogin}, {Kocevski}, {Faber}, {Ferguson},
  {Koekemoer}, {Riess}, {Acquaviva}, {Alexander}, {Almaini}, {Ashby}, {Barden},
  {Bell}, {Bournaud}, {Brown}, {Caputi}, {Casertano}, {Cassata}, {Castellano},
  {Challis}, {Chary}, {Cheung}, {Cirasuolo}, {Conselice}, {Roshan Cooray},
  {Croton}, {Daddi}, {Dahlen}, {Dav{\'e}}, {de Mello}, {Dekel}, {Dickinson},
  {Dolch}, {Donley}, {Dunlop}, {Dutton}, {Elbaz}, {Fazio}, {Filippenko},
  {Finkelstein}, {Fontana}, {Gardner}, {Garnavich}, {Gawiser}, {Giavalisco},
  {Grazian}, {Guo}, {Hathi}, {H{\"a}ussler}, {Hopkins}, {Huang}, {Huang},
  {Jha}, {Kartaltepe}, {Kirshner}, {Koo}, {Lai}, {Lee}, {Li}, {Lotz}, {Lucas},
  {Madau}, {McCarthy}, {McGrath}, {McIntosh}, {McLure}, {Mobasher},
  {Moustakas}, {Mozena}, {Nandra}, {Newman}, {Niemi}, {Noeske}, {Papovich},
  {Pentericci}, {Pope}, {Primack}, {Rajan}, {Ravindranath}, {Reddy}, {Renzini},
  {Rix}, {Robaina}, {Rodney}, {Rosario}, {Rosati}, {Salimbeni}, {Scarlata},
  {Siana}, {Simard}, {Smidt}, {Somerville}, {Spinrad}, {Straughn}, {Strolger},
  {Telford}, {Teplitz}, {Trump}, {van der Wel}, {Villforth}, {Wechsler},
  {Weiner}, {Wiklind}, {Wild}, {Wilson}, {Wuyts}, {Yan}, \& {Yun}}]{Grogin2011}
{Grogin}, N.~A., {et~al.} 2011, \apjs, 197, 35

\bibitem[{{Guo} {et~al.}(2011){Guo}, {White}, {Boylan-Kolchin}, {De Lucia},
  {Kauffmann}, {Lemson}, {Li}, {Springel}, \& {Weinmann}}]{Guo2011}
{Guo}, Q., {et~al.} 2011, \mnras, 413, 101

\bibitem[{{Hasinger} {et~al.}(2007){Hasinger}, {Cappelluti}, {Brunner},
  {Brusa}, {Comastri}, {Elvis}, {Finoguenov}, {Fiore}, {Franceschini}, {Gilli},
  {Griffiths}, {Lehmann}, {Mainieri}, {Matt}, {Matute}, {Miyaji}, {Molendi},
  {Paltani}, {Sanders}, {Scoville}, {Tresse}, {Urry}, {Vettolani}, \&
  {Zamorani}}]{Hasinger2007}
{Hasinger}, G., {et~al.} 2007, \apjs, 172, 29

\bibitem[{{Henriques} {et~al.}(2012){Henriques}, {White}, {Lemson}, {Thomas},
  {Guo}, {Marleau}, \& {Overzier}}]{Henriques2012}
{Henriques}, B.~M.~B., {White}, S.~D.~M., {Lemson}, G., {Thomas}, P.~A., {Guo},
  Q., {Marleau}, G.-D., \& {Overzier}, R.~A. 2012, \mnras, 421, 2904

\bibitem[{{Ilbert} {et~al.}(2006){Ilbert}, {Arnouts}, {McCracken},
  {Bolzonella}, {Bertin}, {Le F{\`e}vre}, {Mellier}, {Zamorani}, {Pell{\`o}},
  {Iovino}, {Tresse}, {Le Brun}, {Bottini}, {Garilli}, {Maccagni}, {Picat},
  {Scaramella}, {Scodeggio}, {Vettolani}, {Zanichelli}, {Adami}, {Bardelli},
  {Cappi}, {Charlot}, {Ciliegi}, {Contini}, {Cucciati}, {Foucaud}, {Franzetti},
  {Gavignaud}, {Guzzo}, {Marano}, {Marinoni}, {Mazure}, {Meneux}, {Merighi},
  {Paltani}, {Pollo}, {Pozzetti}, {Radovich}, {Zucca}, {Bondi}, {Bongiorno},
  {Busarello}, {de La Torre}, {Gregorini}, {Lamareille}, {Mathez}, {Merluzzi},
  {Ripepi}, {Rizzo}, \& {Vergani}}]{Ilbert2006}
{Ilbert}, O., {et~al.} 2006, \aap, 457, 841

\bibitem[{{Ilbert} {et~al.}(2009){Ilbert}, {Capak}, {Salvato}, {Aussel},
  {McCracken}, {Sanders}, {Scoville}, {Kartaltepe}, {Arnouts}, {Le Floc'h},
  {Mobasher}, {Taniguchi}, {Lamareille}, {Leauthaud}, {Sasaki}, {Thompson},
  {Zamojski}, {Zamorani}, {Bardelli}, {Bolzonella}, {Bongiorno}, {Brusa},
  {Caputi}, {Carollo}, {Contini}, {Cook}, {Coppa}, {Cucciati}, {de la Torre},
  {de Ravel}, {Franzetti}, {Garilli}, {Hasinger}, {Iovino}, {Kampczyk},
  {Kneib}, {Knobel}, {Kovac}, {Le Borgne}, {Le Brun}, {F{\`e}vre}, {Lilly},
  {Looper}, {Maier}, {Mainieri}, {Mellier}, {Mignoli}, {Murayama}, {Pell{\`o}},
  {Peng}, {P{\'e}rez-Montero}, {Renzini}, {Ricciardelli}, {Schiminovich},
  {Scodeggio}, {Shioya}, {Silverman}, {Surace}, {Tanaka}, {Tasca}, {Tresse},
  {Vergani}, \& {Zucca}}]{Ilbert2009}
---. 2009, \apj, 690, 1236

\bibitem[{{Ilbert} {et~al.}(2010){Ilbert}, {Salvato}, {Le Floc'h}, {Aussel},
  {Capak}, {McCracken}, {Mobasher}, {Kartaltepe}, {Scoville}, {Sanders},
  {Arnouts}, {Bundy}, {Cassata}, {Kneib}, {Koekemoer}, {Le F{\`e}vre}, {Lilly},
  {Surace}, {Taniguchi}, {Tasca}, {Thompson}, {Tresse}, {Zamojski}, {Zamorani},
  \& {Zucca}}]{Ilbert2010}
---. 2010, \apj, 709, 644

\bibitem[{{Kennicutt}(1998)}]{Kennicutt1998}
{Kennicutt}, Jr., R.~C. 1998, \araa, 36, 189

\bibitem[{{Koekemoer} {et~al.}(2007){Koekemoer}, {Aussel}, {Calzetti}, {Capak},
  {Giavalisco}, {Kneib}, {Leauthaud}, {Le F{\`e}vre}, {McCracken}, {Massey},
  {Mobasher}, {Rhodes}, {Scoville}, \& {Shopbell}}]{Koekemoer2007}
{Koekemoer}, A.~M., {et~al.} 2007, \apjs, 172, 196

\bibitem[{{Koekemoer} {et~al.}(2011){Koekemoer}, {Faber}, {Ferguson}, {Grogin},
  {Kocevski}, {Koo}, {Lai}, {Lotz}, {Lucas}, {McGrath}, {Ogaz}, {Rajan},
  {Riess}, {Rodney}, {Strolger}, {Casertano}, {Castellano}, {Dahlen},
  {Dickinson}, {Dolch}, {Fontana}, {Giavalisco}, {Grazian}, {Guo}, {Hathi},
  {Huang}, {van der Wel}, {Yan}, {Acquaviva}, {Alexander}, {Almaini}, {Ashby},
  {Barden}, {Bell}, {Bournaud}, {Brown}, {Caputi}, {Cassata}, {Challis},
  {Chary}, {Cheung}, {Cirasuolo}, {Conselice}, {Roshan Cooray}, {Croton},
  {Daddi}, {Dav{\'e}}, {de Mello}, {de Ravel}, {Dekel}, {Donley}, {Dunlop},
  {Dutton}, {Elbaz}, {Fazio}, {Filippenko}, {Finkelstein}, {Frazer}, {Gardner},
  {Garnavich}, {Gawiser}, {Gruetzbauch}, {Hartley}, {H{\"a}ussler},
  {Herrington}, {Hopkins}, {Huang}, {Jha}, {Johnson}, {Kartaltepe},
  {Khostovan}, {Kirshner}, {Lani}, {Lee}, {Li}, {Madau}, {McCarthy},
  {McIntosh}, {McLure}, {McPartland}, {Mobasher}, {Moreira}, {Mortlock},
  {Moustakas}, {Mozena}, {Nandra}, {Newman}, {Nielsen}, {Niemi}, {Noeske},
  {Papovich}, {Pentericci}, {Pope}, {Primack}, {Ravindranath}, {Reddy},
  {Renzini}, {Rix}, {Robaina}, {Rosario}, {Rosati}, {Salimbeni}, {Scarlata},
  {Siana}, {Simard}, {Smidt}, {Snyder}, {Somerville}, {Spinrad}, {Straughn},
  {Telford}, {Teplitz}, {Trump}, {Vargas}, {Villforth}, {Wagner}, {Wandro},
  {Wechsler}, {Weiner}, {Wiklind}, {Wild}, {Wilson}, {Wuyts}, \&
  {Yun}}]{Koekemoer2011}
---. 2011, \apjs, 197, 36

\bibitem[{{Kriek} {et~al.}(2009){Kriek}, {van Dokkum}, {Labb{\'e}}, {Franx},
  {Illingworth}, {Marchesini}, \& {Quadri}}]{Kriek2009}
{Kriek}, M., {van Dokkum}, P.~G., {Labb{\'e}}, I., {Franx}, M., {Illingworth},
  G.~D., {Marchesini}, D., \& {Quadri}, R.~F. 2009, \apj, 700, 221

\bibitem[{{Kriek} {et~al.}(2010){Kriek}, {Labb{\'e}}, {Conroy}, {Whitaker},
  {van Dokkum}, {Brammer}, {Franx}, {Illingworth}, {Marchesini}, {Muzzin},
  {Quadri}, \& {Rudnick}}]{Kriek2010}
{Kriek}, M., {et~al.} 2010, \apjl, 722, L64

\bibitem[{{Kron}(1980)}]{Kron1980}
{Kron}, R.~G. 1980, \apjs, 43, 305

\bibitem[{{Labb{\'e}} {et~al.}(2003){Labb{\'e}}, {Franx}, {Rudnick},
  {Schreiber}, {Rix}, {Moorwood}, {van Dokkum}, {van der Werf},
  {R{\"o}ttgering}, {van Starkenburg}, {van der Wel}, {Kuijken}, \&
  {Daddi}}]{Labbe2003}
{Labb{\'e}}, I., {et~al.} 2003, \aj, 125, 1107

\bibitem[{{Labb{\'e}} {et~al.}(2005){Labb{\'e}}, {Huang}, {Franx}, {Rudnick},
  {Barmby}, {Daddi}, {van Dokkum}, {Fazio}, {Schreiber}, {Moorwood}, {Rix},
  {R{\"o}ttgering}, {Trujillo}, \& {van der Werf}}]{Labbe2005}
---. 2005, \apjl, 624, L81

\bibitem[{{Labb{\'e}} {et~al.}(2010){Labb{\'e}}, {Gonz{\'a}lez}, {Bouwens},
  {Illingworth}, {Franx}, {Trenti}, {Oesch}, {van Dokkum}, {Stiavelli},
  {Carollo}, {Kriek}, \& {Magee}}]{Labbe2010}
---. 2010, \apjl, 716, L103

\bibitem[{{Labbe} {et~al.}(2012){Labbe}, {Oesch}, {Bouwens}, {Illingworth},
  {Magee}, {Gonzalez}, {Carollo}, {Franx}, {Trenti}, {van Dokkum}, \&
  {Stiavelli}}]{Labbe2012}
{Labbe}, I., {et~al.} 2012, ArXiv e-prints

\bibitem[{{Lawrence} {et~al.}(2007){Lawrence}, {Warren}, {Almaini}, {Edge},
  {Hambly}, {Jameson}, {Lucas}, {Casali}, {Adamson}, {Dye}, {Emerson},
  {Foucaud}, {Hewett}, {Hirst}, {Hodgkin}, {Irwin}, {Lodieu}, {McMahon},
  {Simpson}, {Smail}, {Mortlock}, \& {Folger}}]{Lawrence2007}
{Lawrence}, A., {et~al.} 2007, \mnras, 379, 1599

\bibitem[{{Lilly} {et~al.}(2007){Lilly}, {Le F{\`e}vre}, {Renzini}, {Zamorani},
  {Scodeggio}, {Contini}, {Carollo}, {Hasinger}, {Kneib}, {Iovino}, {Le Brun},
  {Maier}, {Mainieri}, {Mignoli}, {Silverman}, {Tasca}, {Bolzonella},
  {Bongiorno}, {Bottini}, {Capak}, {Caputi}, {Cimatti}, {Cucciati}, {Daddi},
  {Feldmann}, {Franzetti}, {Garilli}, {Guzzo}, {Ilbert}, {Kampczyk}, {Kovac},
  {Lamareille}, {Leauthaud}, {Borgne}, {McCracken}, {Marinoni}, {Pello},
  {Ricciardelli}, {Scarlata}, {Vergani}, {Sanders}, {Schinnerer}, {Scoville},
  {Taniguchi}, {Arnouts}, {Aussel}, {Bardelli}, {Brusa}, {Cappi}, {Ciliegi},
  {Finoguenov}, {Foucaud}, {Franceschini}, {Halliday}, {Impey}, {Knobel},
  {Koekemoer}, {Kurk}, {Maccagni}, {Maddox}, {Marano}, {Marconi}, {Meneux},
  {Mobasher}, {Moreau}, {Peacock}, {Porciani}, {Pozzetti}, {Scaramella},
  {Schiminovich}, {Shopbell}, {Smail}, {Thompson}, {Tresse}, {Vettolani},
  {Zanichelli}, \& {Zucca}}]{Lilly2007}
{Lilly}, S.~J., {et~al.} 2007, \apjs, 172, 70

\bibitem[{{Lilly} {et~al.}(2009){Lilly}, {Le Brun}, {Maier}, {Mainieri},
  {Mignoli}, {Scodeggio}, {Zamorani}, {Carollo}, {Contini}, {Kneib}, {Le
  F{\`e}vre}, {Renzini}, {Bardelli}, {Bolzonella}, {Bongiorno}, {Caputi},
  {Coppa}, {Cucciati}, {de la Torre}, {de Ravel}, {Franzetti}, {Garilli},
  {Iovino}, {Kampczyk}, {Kovac}, {Knobel}, {Lamareille}, {Le Borgne}, {Pello},
  {Peng}, {P{\'e}rez-Montero}, {Ricciardelli}, {Silverman}, {Tanaka}, {Tasca},
  {Tresse}, {Vergani}, {Zucca}, {Ilbert}, {Salvato}, {Oesch}, {Abbas},
  {Bottini}, {Capak}, {Cappi}, {Cassata}, {Cimatti}, {Elvis}, {Fumana},
  {Guzzo}, {Hasinger}, {Koekemoer}, {Leauthaud}, {Maccagni}, {Marinoni},
  {McCracken}, {Memeo}, {Meneux}, {Porciani}, {Pozzetti}, {Sanders},
  {Scaramella}, {Scarlata}, {Scoville}, {Shopbell}, \& {Taniguchi}}]{Lilly2009}
---. 2009, \apjs, 184, 218

\bibitem[{{Ma{\'{\i}}z Apell{\'a}niz}(2006)}]{MaizApellaniz2006}
{Ma{\'{\i}}z Apell{\'a}niz}, J. 2006, \aj, 131, 1184

\bibitem[{{Maraston}(2005)}]{Maraston2005}
{Maraston}, C. 2005, \mnras, 362, 799

\bibitem[{{Maraston} {et~al.}(2010){Maraston}, {Pforr}, {Renzini}, {Daddi},
  {Dickinson}, {Cimatti}, \& {Tonini}}]{Maraston2010}
{Maraston}, C., {Pforr}, J., {Renzini}, A., {Daddi}, E., {Dickinson}, M.,
  {Cimatti}, A., \& {Tonini}, C. 2010, \mnras, 407, 830

\bibitem[{{Marchesini} {et~al.}(2012){Marchesini}, {Stefanon}, {Brammer}, \&
  {Whitaker}}]{Marchesini2012}
{Marchesini}, D., {Stefanon}, M., {Brammer}, G.~B., \& {Whitaker}, K.~E. 2012,
  \apj, 748, 126

\bibitem[{{Marchesini} {et~al.}(2009){Marchesini}, {van Dokkum}, {F{\"o}rster
  Schreiber}, {Franx}, {Labb{\'e}}, \& {Wuyts}}]{Marchesini2009}
{Marchesini}, D., {van Dokkum}, P.~G., {F{\"o}rster Schreiber}, N.~M., {Franx},
  M., {Labb{\'e}}, I., \& {Wuyts}, S. 2009, \apj, 701, 1765

\bibitem[{{Marchesini} {et~al.}(2010){Marchesini}, {Whitaker}, {Brammer}, {van
  Dokkum}, {Labb{\'e}}, {Muzzin}, {Quadri}, {Kriek}, {Lee}, {Rudnick}, {Franx},
  {Illingworth}, \& {Wake}}]{Marchesini2010}
{Marchesini}, D., {et~al.} 2010, \apj, 725, 1277

\bibitem[{{Martin} {et~al.}(2005){Martin}, {Fanson}, {Schiminovich},
  {Morrissey}, {Friedman}, {Barlow}, {Conrow}, {Grange}, {Jelinsky},
  {Milliard}, {Siegmund}, {Bianchi}, {Byun}, {Donas}, {Forster}, {Heckman},
  {Lee}, {Madore}, {Malina}, {Neff}, {Rich}, {Small}, {Surber}, {Szalay},
  {Welsh}, \& {Wyder}}]{Martin2005}
{Martin}, D.~C., {et~al.} 2005, \apjl, 619, L1

\bibitem[{{McCracken} {et~al.}(2010){McCracken}, {Capak}, {Salvato}, {Aussel},
  {Thompson}, {Daddi}, {Sanders}, {Kneib}, {Willott}, {Mancini}, {Renzini},
  {Cook}, {Le F{\`e}vre}, {Ilbert}, {Kartaltepe}, {Koekemoer}, {Mellier},
  {Murayama}, {Scoville}, {Shioya}, \& {Tanaguchi}}]{McCracken2010}
{McCracken}, H.~J., {et~al.} 2010, \apj, 708, 202

\bibitem[{{McCracken} {et~al.}(2012){McCracken}, {Milvang-Jensen}, {Dunlop},
  {Franx}, {Fynbo}, {Le F{\`e}vre}, {Holt}, {Caputi}, {Goranova}, {Buitrago},
  {Emerson}, {Freudling}, {Hudelot}, {L{\'o}pez-Sanjuan}, {Magnard}, {Mellier},
  {M{\o}ller}, {Nilsson}, {Sutherland}, {Tasca}, \& {Zabl}}]{McCracken2012}
---. 2012, \aap, 544, A156

\bibitem[{{McLure} {et~al.}(2006){McLure}, {Cirasuolo}, {Dunlop}, {Sekiguchi},
  {Almaini}, {Foucaud}, {Simpson}, {Watson}, {Hirst}, {Page}, \&
  {Smail}}]{Mclure2006}
{McLure}, R.~J., {et~al.} 2006, \mnras, 372, 357

\bibitem[{{Muzzin} {et~al.}(2010){Muzzin}, {van Dokkum}, {Kriek}, {Labb{\'e}},
  {Cury}, {Marchesini}, \& {Franx}}]{Muzzin2010}
{Muzzin}, A., {van Dokkum}, P., {Kriek}, M., {Labb{\'e}}, I., {Cury}, I.,
  {Marchesini}, D., \& {Franx}, M. 2010, \apj, 725, 742

\bibitem[{{Muzzin} {et~al.}(2013){Muzzin}, {Marchesini}, {Stefanon}, {Franx},
  {McCracken}, {Milvang-Jensen}, {Dunlop}, {Fynbo}, {Le Fevre}, {Brammer}, \&
  {Labbe}}]{Muzzin2013c}
{Muzzin}, A., {et~al.} 2013, arXiv:1303.4409

\bibitem[{{Noeske} {et~al.}(2007){Noeske}, {Weiner}, {Faber}, {Papovich},
  {Koo}, {Somerville}, {Bundy}, {Conselice}, {Newman}, {Schiminovich}, {Le
  Floc'h}, {Coil}, {Rieke}, {Lotz}, {Primack}, {Barmby}, {Cooper}, {Davis},
  {Ellis}, {Fazio}, {Guhathakurta}, {Huang}, {Kassin}, {Martin}, {Phillips},
  {Rich}, {Small}, {Willmer}, \& {Wilson}}]{Noeske2007}
{Noeske}, K.~G., {et~al.} 2007, \apjl, 660, L43

\bibitem[{{Oliver} {et~al.}(2012){Oliver}, {Bock}, {Altieri}, {Amblard},
  {Arumugam}, {Aussel}, {Babbedge}, {Beelen}, {B{\'e}thermin}, {Blain},
  {Boselli}, {Bridge}, {Brisbin}, {Buat}, {Burgarella},
  {Castro-Rodr{\'{\i}}guez}, {Cava}, {Chanial}, {Cirasuolo}, {Clements},
  {Conley}, {Conversi}, {Cooray}, {Dowell}, {Dubois}, {Dwek}, {Dye}, {Eales},
  {Elbaz}, {Farrah}, {Feltre}, {Ferrero}, {Fiolet}, {Fox}, {Franceschini},
  {Gear}, {Giovannoli}, {Glenn}, {Gong}, {Gonz{\'a}lez Solares}, {Griffin},
  {Halpern}, {Harwit}, {Hatziminaoglou}, {Heinis}, {Hurley}, {Hwang}, {Hyde},
  {Ibar}, {Ilbert}, {Isaak}, {Ivison}, {Lagache}, {Le Floc'h}, {Levenson},
  {Faro}, {Lu}, {Madden}, {Maffei}, {Magdis}, {Mainetti}, {Marchetti},
  {Marsden}, {Marshall}, {Mortier}, {Nguyen}, {O'Halloran}, {Omont}, {Page},
  {Panuzzo}, {Papageorgiou}, {Patel}, {Pearson}, {P{\'e}rez-Fournon}, {Pohlen},
  {Rawlings}, {Raymond}, {Rigopoulou}, {Riguccini}, {Rizzo}, {Rodighiero},
  {Roseboom}, {Rowan-Robinson}, {S{\'a}nchez Portal}, {Schulz}, {Scott},
  {Seymour}, {Shupe}, {Smith}, {Stevens}, {Symeonidis}, {Trichas}, {Tugwell},
  {Vaccari}, {Valtchanov}, {Vieira}, {Viero}, {Vigroux}, {Wang}, {Ward},
  {Wardlow}, {Wright}, {Xu}, \& {Zemcov}}]{Oliver2012}
{Oliver}, S.~J., {et~al.} 2012, \mnras, 424, 1614

\bibitem[{{Onodera} {et~al.}(2012){Onodera}, {Renzini}, {Carollo},
  {Cappellari}, {Mancini}, {Strazzullo}, {Daddi}, {Arimoto}, {Gobat}, {Yamada},
  {McCracken}, {Ilbert}, {Capak}, {Cimatti}, {Giavalisco}, {Koekemoer}, {Kong},
  {Lilly}, {Motohara}, {Ohta}, {Sanders}, {Scoville}, {Tamura}, \&
  {Taniguchi}}]{Onodera2012}
{Onodera}, M., {et~al.} 2012, \apj, 755, 26

\bibitem[{{Papovich} {et~al.}(2011){Papovich}, {Finkelstein}, {Ferguson},
  {Lotz}, \& {Giavalisco}}]{Papovich2011}
{Papovich}, C., {Finkelstein}, S.~L., {Ferguson}, H.~C., {Lotz}, J.~M., \&
  {Giavalisco}, M. 2011, \mnras, 412, 1123

\bibitem[{{Patel} {et~al.}(2012{\natexlab{a}}){Patel}, {Holden}, {Kelson},
  {Franx}, {van der Wel}, \& {Illingworth}}]{Patel2012a}
{Patel}, S.~G., {Holden}, B.~P., {Kelson}, D.~D., {Franx}, M., {van der Wel},
  A., \& {Illingworth}, G.~D. 2012{\natexlab{a}}, \apjl, 748, L27

\bibitem[{{Patel} {et~al.}(2012{\natexlab{b}}){Patel}, {van Dokkum}, {Franx},
  {Quadri}, {Muzzin}, {Marchesini}, {Williams}, {Holden}, \&
  {Stefanon}}]{Patel2012b}
{Patel}, S.~G., {et~al.} 2012{\natexlab{b}}, ArXiv e-prints

\bibitem[{{Quadri} {et~al.}(2007){Quadri}, {van Dokkum}, {Gawiser}, {Franx},
  {Marchesini}, {Lira}, {Rudnick}, {Herrera}, {Maza}, {Kriek}, {Labb{\'e}}, \&
  {Francke}}]{Quadri2007}
{Quadri}, R., {et~al.} 2007, \apj, 654, 138

\bibitem[{{Sanders} {et~al.}(2007){Sanders}, {Salvato}, {Aussel}, {Ilbert},
  {Scoville}, {Surace}, {Frayer}, {Sheth}, {Helou}, {Brooke}, {Bhattacharya},
  {Yan}, {Kartaltepe}, {Barnes}, {Blain}, {Calzetti}, {Capak}, {Carilli},
  {Carollo}, {Comastri}, {Daddi}, {Ellis}, {Elvis}, {Fall}, {Franceschini},
  {Giavalisco}, {Hasinger}, {Impey}, {Koekemoer}, {Le F{\`e}vre}, {Lilly},
  {Liu}, {McCracken}, {Mobasher}, {Renzini}, {Rich}, {Schinnerer}, {Shopbell},
  {Taniguchi}, {Thompson}, {Urry}, \& {Williams}}]{Sanders2007}
{Sanders}, D.~B., {et~al.} 2007, \apjs, 172, 86

\bibitem[{{Schinnerer} {et~al.}(2007){Schinnerer}, {Smol{\v c}i{\'c}},
  {Carilli}, {Bondi}, {Ciliegi}, {Jahnke}, {Scoville}, {Aussel}, {Bertoldi},
  {Blain}, {Impey}, {Koekemoer}, {Le Fevre}, \& {Urry}}]{Shinnerer2007}
{Schinnerer}, E., {et~al.} 2007, \apjs, 172, 46

\bibitem[{{Schinnerer} {et~al.}(2010){Schinnerer}, {Sargent}, {Bondi}, {Smol{\v
  c}i{\'c}}, {Datta}, {Carilli}, {Bertoldi}, {Blain}, {Ciliegi}, {Koekemoer},
  \& {Scoville}}]{Shinnerer2010}
---. 2010, \apjs, 188, 384

\bibitem[{{Schlegel} {et~al.}(1998){Schlegel}, {Finkbeiner}, \&
  {Davis}}]{Schlegel1998}
{Schlegel}, D.~J., {Finkbeiner}, D.~P., \& {Davis}, M. 1998, \apj, 500, 525

\bibitem[{{Scott} {et~al.}(2008){Scott}, {Austermann}, {Perera}, {Wilson},
  {Aretxaga}, {Bock}, {Hughes}, {Kang}, {Kim}, {Mauskopf}, {Sanders},
  {Scoville}, \& {Yun}}]{Scott2008}
{Scott}, K.~S., {et~al.} 2008, \mnras, 385, 2225

\bibitem[{{Scoville} {et~al.}(2007){Scoville}, {Aussel}, {Brusa}, {Capak},
  {Carollo}, {Elvis}, {Giavalisco}, {Guzzo}, {Hasinger}, {Impey}, {Kneib},
  {LeFevre}, {Lilly}, {Mobasher}, {Renzini}, {Rich}, {Sanders}, {Schinnerer},
  {Schminovich}, {Shopbell}, {Taniguchi}, \& {Tyson}}]{Scoville2007}
{Scoville}, N., {et~al.} 2007, \apjs, 172, 1

\bibitem[{{Skrutskie} {et~al.}(2006){Skrutskie}, {Cutri}, {Stiening},
  {Weinberg}, {Schneider}, {Carpenter}, {Beichman}, {Capps}, {Chester},
  {Elias}, {Huchra}, {Liebert}, {Lonsdale}, {Monet}, {Price}, {Seitzer},
  {Jarrett}, {Kirkpatrick}, {Gizis}, {Howard}, {Evans}, {Fowler}, {Fullmer},
  {Hurt}, {Light}, {Kopan}, {Marsh}, {McCallon}, {Tam}, {Van Dyk}, \&
  {Wheelock}}]{Skrutskie2006}
{Skrutskie}, M.~F., {et~al.} 2006, \aj, 131, 1163

\bibitem[{{Taniguchi} {et~al.}(2007){Taniguchi}, {Scoville}, {Murayama},
  {Sanders}, {Mobasher}, {Aussel}, {Capak}, {Ajiki}, {Miyazaki}, {Komiyama},
  {Shioya}, {Nagao}, {Sasaki}, {Koda}, {Carilli}, {Giavalisco}, {Guzzo},
  {Hasinger}, {Impey}, {LeFevre}, {Lilly}, {Renzini}, {Rich}, {Schinnerer},
  {Shopbell}, {Kaifu}, {Karoji}, {Arimoto}, {Okamura}, \&
  {Ohta}}]{Taniguchi2007}
{Taniguchi}, Y., {et~al.} 2007, \apjs, 172, 9

\bibitem[{{van de Sande} {et~al.}(2011){van de Sande}, {Kriek}, {Franx}, {van
  Dokkum}, {Bezanson}, {Whitaker}, {Brammer}, {Labb{\'e}}, {Groot}, \&
  {Kaper}}]{vandesande2011}
{van de Sande}, J., {et~al.} 2011, \apjl, 736, L9

\bibitem[{{van de Sande} {et~al.}(2012){van de Sande}, {Kriek}, {Franx}, {van
  Dokkum}, {Bezanson}, {Bouwens}, {Quadri}, {Rix}, \&
  {Skelton}}]{vandesande2012}
---. 2012, ArXiv e-prints

\bibitem[{{van Dokkum} {et~al.}(2006){van Dokkum}, {Quadri}, {Marchesini},
  {Rudnick}, {Franx}, {Gawiser}, {Herrera}, {Wuyts}, {Lira}, {Labb{\'e}},
  {Maza}, {Illingworth}, {F{\"o}rster Schreiber}, {Kriek}, {Rix}, {Taylor},
  {Toft}, {Webb}, \& {Yi}}]{vanDokkum2006}
{van Dokkum}, P.~G., {et~al.} 2006, \apjl, 638, L59

\bibitem[{{van Dokkum} {et~al.}(2009){van Dokkum}, {Labb{\'e}}, {Marchesini},
  {Quadri}, {Brammer}, {Whitaker}, {Kriek}, {Franx}, {Rudnick}, {Illingworth},
  {Lee}, \& {Muzzin}}]{vanDokkum2009b}
---. 2009, \pasp, 121, 2

\bibitem[{{Whitaker} {et~al.}(2012{\natexlab{a}}){Whitaker}, {Kriek}, {van
  Dokkum}, {Bezanson}, {Brammer}, {Franx}, \& {Labb{\'e}}}]{Whitaker2012b}
{Whitaker}, K.~E., {Kriek}, M., {van Dokkum}, P.~G., {Bezanson}, R., {Brammer},
  G., {Franx}, M., \& {Labb{\'e}}, I. 2012{\natexlab{a}}, \apj, 745, 179

\bibitem[{{Whitaker} {et~al.}(2012{\natexlab{b}}){Whitaker}, {van Dokkum},
  {Brammer}, \& {Franx}}]{Whitaker2012}
{Whitaker}, K.~E., {van Dokkum}, P.~G., {Brammer}, G., \& {Franx}, M.
  2012{\natexlab{b}}, \apjl, 754, L29

\bibitem[{{Whitaker} {et~al.}(2011){Whitaker}, {Labb{\'e}}, {van Dokkum},
  {Brammer}, {Kriek}, {Marchesini}, {Quadri}, {Franx}, {Muzzin}, {Williams},
  {Bezanson}, {Illingworth}, {Lee}, {Lundgren}, {Nelson}, {Rudnick}, {Tal}, \&
  {Wake}}]{Whitaker2011}
{Whitaker}, K.~E., {et~al.} 2011, \apj, 735, 86

\bibitem[{{Williams} {et~al.}(2009){Williams}, {Quadri}, {Franx}, {van Dokkum},
  \& {Labb{\'e}}}]{Williams2009}
{Williams}, R.~J., {Quadri}, R.~F., {Franx}, M., {van Dokkum}, P., \&
  {Labb{\'e}}, I. 2009, \apj, 691, 1879

\bibitem[{{Wuyts} {et~al.}(2008){Wuyts}, {Labb{\'e}}, {Schreiber}, {Franx},
  {Rudnick}, {Brammer}, \& {van Dokkum}}]{Wuyts2008}
{Wuyts}, S., {Labb{\'e}}, I., {Schreiber}, N.~M.~F., {Franx}, M., {Rudnick},
  G., {Brammer}, G.~B., \& {van Dokkum}, P.~G. 2008, \apj, 682, 985

\bibitem[{{Wuyts} {et~al.}(2007){Wuyts}, {Labb{\'e}}, {Franx}, {Rudnick}, {van
  Dokkum}, {Fazio}, {F{\"o}rster Schreiber}, {Huang}, {Moorwood}, {Rix},
  {R{\"o}ttgering}, \& {van der Werf}}]{Wuyts2007}
{Wuyts}, S., {et~al.} 2007, \apj, 655, 51

\bibitem[{{Wuyts} {et~al.}(2011){Wuyts}, {F{\"o}rster Schreiber}, {van der
  Wel}, {Magnelli}, {Guo}, {Genzel}, {Lutz}, {Aussel}, {Barro}, {Berta},
  {Cava}, {Graci{\'a}-Carpio}, {Hathi}, {Huang}, {Kocevski}, {Koekemoer},
  {Lee}, {Le Floc'h}, {McGrath}, {Nordon}, {Popesso}, {Pozzi}, {Riguccini},
  {Rodighiero}, {Saintonge}, \& {Tacconi}}]{Wuyts2011}
---. 2011, \apj, 742, 96

\end{thebibliography}
\begin{appendix}
\begin{center}
COMPARISON OF ULTRAVISTA K$_{s}$ ZEROPOINT TO OTHER DATASETS
\end{center}
In this appendix we make a comparison between the K$_{s}$-band photometry between several datasets that cover the COSMOS field to test the photometric zeropoint of the UltraVISTA K$_{s}$ imaging.  For the catalog, the zeropoint in K$_{s}$ is important because K$_{s}$ is used as the anchor filter when calculating the zeropoint offsets in the remaining filters ($\S$ 4.1).  We compare the UltraVISTA K$_{s}$-band to the K-band imaging from the NMBS \citep{Whitaker2011}, which reaches a 5$\sigma$ depth of K = 23.5.  We also compare the CFHT/WIRCAM K$_{s}$ imaging from \cite{McCracken2010}, which in the deepest regions reaches a 5$\sigma$ depth of K$_{s}$ = 23.65 in a 2$^{\prime\prime}$ aperture.  Lastly, we compare to sources detected in 2MASS \citep{Skrutskie2006}, although we note that the overlap in in dynamic range between UltraVISTA and 2MASS is limited.
\newline\indent
For comparison with the WIRCAM and NMBS data we use the region of the survey that is in common to all three datasets.  This region covers $\sim$ 0.22 deg$^2$ and is located in the northwest corner of the UltraVISTA field.  We perform PSF matching on all three datasets, matching to the worst-seeing image, which is the NMBS ($\sim$ 1.1$^{\prime\prime}$).  Ten bright unsaturated stars in the region are used in order to determine the convolution kernel.  Once the PSFs are homogenized we use SExtractor to find objects on each image individually and measure fluxes within a 2.1$^{\prime\prime}$ diameter aperture.  Object detection is performed on each image separately because there are small, but noticable differences in the astrometry between the surveys.  If object detection was performed with SExtractor in dual-image mode, small astrometry differences could appear as zeropoint differences.
Source lists from each catalog are matched using a 0.5$^{\prime\prime}$ search radius.  Using a small radius means that some matches may be missed, but the tradeoff is that there are few ambiguous matches.  
\newline\indent
In Figure 14 we plot the comparison of the UltraVISTA K$_{s}$ photometry vs. the WIRCAM and NMBS K$_{s}$ photometry.  In general, the comparison is good.  There is a small systematic difference between the surveys; however, there is no systematic difference as a function of magnitude.   If we compare the differences in the magnitude range 16 $< K_{s} <$ 20, we find an offset between UltraVISTA and WIRCAM of 0.051 mag, and an offset between UltraVISTA and NMBS of 0.078 mag, with UltraVISTA being fainter in both cases.
\newline\indent
The difference of 0.054 mag between the UltraVISTA and the WIRCAM data is very similar to that measured by \cite{McCracken2012} (see their Figure 10) for BzK-selected stars in the fields of the two surveys.  The difference between UltraVISTA and the NMBS is slightly larger than the difference between UltraVISTA and WIRCAM.  That difference implies that the WIRCAM photometry should be 0.027 mag fainter than the NMBS photometry.  Indeed, this implied difference is very similar to the actual difference measured by \cite{Whitaker2011} who performed PSF matched photometry between the NMBS and the WIRCAM data (see their Figure 12).  
\newline\indent
In the bottom panel of Figure 14 we also compare the UltraVISTA photometry to photometry from the 2MASS catalog.  For this comparison we do not perform PSF matching, but match objects between the surveys over the full 1.62 deg$^2$ UltraVISTA field.  We compare the total magnitudes measured for UltraVISTA (before any zeropoint offset has been applied), to the ``default" magnitudes of objects extracted from the 2MASS point source catalog.  As Figure 14 shows, there is reasonable agreement; however, the dynamic range over which the comparison can be made is limited.  If we compare objects at 15.5 $< K_{s} <$ 17.0 we find a systematic difference of 0.054 mag, with again UltraVISTA being slightly fainter.  
\newline\indent
Although it is not completely clear which zeropoint is the most trustworthy, these comparisons do suggest that the DR1 UltraVISTA K$_{s}$ band zeropoint may be too large at the 0.05 -- 0.08 mag level.  In order to provide consistency with our previous work, we have elected to adjust the zeropoint, making it 0.08 mag brighter so that it matches the NMBS zeropoint.  We note that this offset in zeropoint does not have a significant impact on the colors in the catalog because the zeropoint offsets in the other bands are derived relative to the K$_{s}$-band.  This means that parameters derived from the colors such as the $z_{phot}$ are also unchanged.  The best-fit M/L ratios are also unchanged but the scaling of the K$_{s}$ band means that the M$_{star}$ are systematically changed by adjusting the K$_{s}$ zeropoint; however, this change is quite small.  The offset of 0.08 mag corresponds to a difference of 0.03 dex in M$_{star}$.  If we were to apply the zeropoint implied by 2MASS or the WIRCAM data, it would imply stellar masses that are only 0.01 dex different than those in the current catalog.
\begin{figure*}
\plotone{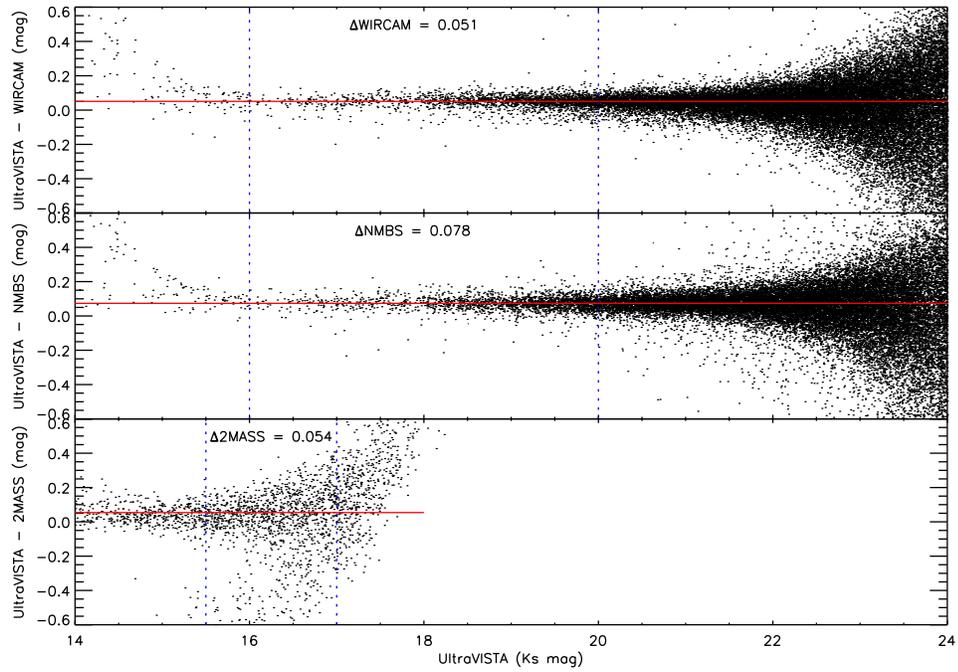}
\caption{\footnotesize Top two panels: Comparison between K$_{s}$ photometry in UltraVISTA to the K$_{s}$ and K photometry from WIRCAM and the NMBS.  Magnitudes have been measured in a 2.1$^{\prime\prime}$ aperture on PSF matched images.  Bottom Panel: Comparison of the UltraVISTA K$_{s}$ photometry to 2MASS K$_{s}$ photometry over the full UltraVISTA area.  Blue dotted lines indicate the magnitude range used to compute the median offset, and the red solid line shows the median offset.  There is a $<$ 0.1 mag difference between UltraVISTA and the other surveys; however, UltraVISTA is systematically too faint and so we have elected to redefine the K$_{s}$ zeropoint to match the NMBS zeropoint.}
\end{figure*}
\end{appendix}




\end{document}